\pdfoutput=1

\documentclass[11pt,nofootinbib]{article}
\usepackage{amsmath,amssymb,mathtools}
\usepackage{graphicx}
\usepackage[colorlinks=true,linkcolor=blue,citecolor=blue,urlcolor=blue,linktocpage=true]{hyperref}
\usepackage{subfigure}
\usepackage{comment}
\usepackage{tensor}
\usepackage{cases}
\usepackage[utf8]{inputenc}
\usepackage{cite}
\usepackage[font=small,labelfont=bf]{caption}

\numberwithin{equation}{section}

\newcommand{\Hinf}{H_{\rm inf}}

\begin{document}

\providecommand{\abs}[1]{\lvert#1\rvert}
\providecommand{\bd}[1]{\boldsymbol{#1}}

\begin{titlepage}

\setcounter{page}{1} \baselineskip=15.5pt \thispagestyle{empty}

\begin{flushright}
\end{flushright}
\vfil

\bigskip
\begin{center}
 {\LARGE \textbf{Reheating-Induced Axion Dark Matter}}\\
 \medskip
 {\LARGE \textbf{After Low Scale Inflation}}
\vskip 15pt
\end{center}

\vspace{0.5cm}
\begin{center}
{\large
Takeshi Kobayashi$^{\star}$
and
Lorenzo Ubaldi$^{\ast,\dagger}$
}\end{center}

\vspace{0.3cm}

\begin{center}
\textit{$^{\star}$ Kobayashi-Maskawa Institute for the Origin of Particles and the Universe,\\ Nagoya University, Nagoya 464-8602, Japan}\\

\vskip 14pt 
\textit{$^{\ast}$ SISSA and INFN Sezione di Trieste, Via Bonomea 265, 34136 Trieste, Italy}\\

\vskip 14pt 
\textit{$^{\dagger}$ Institute for Fundamental Physics of the Universe,
 Via Beirut 2, 34014 Trieste, Italy}\\
 
\vskip 14pt
E-mail:
 \texttt{\href{mailto:takeshi@kmi.nagoya-u.ac.jp}{takeshi@kmi.nagoya-u.ac.jp}},
 \texttt{\href{mailto:ubaldi.physics@gmail.com}{ubaldi.physics@gmail.com}}
\end{center} 



\vspace{1cm}

\noindent
A kinetic mixing between the axion and the inflaton allows for a production of axion dark matter even if the inflationary Hubble scale is smaller than the zero-temperature axion mass. We analyze the axion dynamics in this recently discovered ``inflaxion'' framework, and present a new cosmological scenario where the axion drifts away from its vacuum during the reheating epoch, giving rise to the observed dark matter abundance. We discuss the implications for both the QCD axion and axion-like particles.
\vfil

\end{titlepage}

\newpage
\tableofcontents

\section{Introduction}
\label{sec:intro}

Most of the matter in the Universe is dark, but we have yet to discover what it is composed of. 
A plausible candidate is an axion, a spin zero boson which appears in many models of particle physics.
The best motivated one is the QCD axion~\cite{Weinberg:1977ma, Wilczek:1977pj}, which arises
as a solution to the strong CP problem~\cite{Peccei:1977hh}. 
String theory compactifications also provide motivation to 
contemplate axions, whose masses and couplings span many
orders of magnitude~\cite{Svrcek:2006yi,Arvanitaki:2009fg}.
There is a growing 
experimental effort aimed at covering much of this parameter space, with the hope of detecting a dark matter 
axion. On the other hand, there is still room on the theoretical side to explore new production mechanisms 
for axion dark matter, that can open up new regions of parameters, and
provide guidance for the experiments. 
The aim of this work is to present a new production scenario, by
building up and expanding on an idea we have recently put forward~\cite{Kobayashi:2019eyg}.

An axion is a pseudo Nambu-Goldstone Boson 
of a spontaneously broken U(1) global symmetry (which is the 
Peccei-Quinn symmetry~\cite{Peccei:1977hh} for the QCD axion),
and is characterized by the scale of symmetry breaking~$f$.
If the U(1) is also explicitly broken by a coupling with a strong gauge
group, as is the case for the QCD axion, 
then the confinement scale~$\Lambda$ 
is another important quantity. 
Given a hierarchy between the two scales $\Lambda \ll f$, the
zero-temperature axion mass $m \sim \Lambda^2 / f$ is suppressed.
In textbook treatments and in a large part of the literature it is usually assumed that the 
Hubble scale of inflation is high compared to the axion mass, 
\begin{equation}
 \Hinf \gg m.
\end{equation}
Then, an important distinction is whether the maximum value between $\Hinf$ and the largest temperature~$T_{\mathrm{max}}$
reached in the Universe, is larger or smaller than $f$. 
If it is larger,
$\mathrm{max.} \left\{ H_{\mathrm{inf}}, T_{\mathrm{max}} \right\}
\gg f$, 
the Universe goes through a phase transition from
unbroken to broken U(1) after inflation. 
This gives rise to topological defects including axionic
strings, which later emit axions and give a contribution to the dark
matter relic density, although the actual amount produced is still
uncertain \cite{Gorghetto:2018myk,Kawasaki:2018bzv}. 
If instead 
$\mathrm{max.} \left\{ H_{\mathrm{inf}}, T_{\mathrm{max}}
\right\} \ll f$, 
the U(1) is broken during and after inflation. 
Then the main source of axion dark matter production is the random
initial displacement of the axion field from the vacuum, and this is
often referred to as the vacuum misalignment scenario
\cite{Preskill:1982cy,Abbott:1982af,Dine:1982ah}.

However, we still do not know the scale of inflation. In fact, it could be 
as low as 
\begin{equation}
 \Hinf < m , 
\label{eq:1.1}
\end{equation}
as long as there is enough energy density available
to reheat the Universe to a temperature above MeV where Big Bang
Nucleosynthesis (BBN) takes place. 
With such a low $\Hinf$,
the axion is considered to undergo damped oscillations during inflation and settles
at the minimum of its potential, reaching the point of zero energy density and thus contributing nothing to the dark 
matter abundance. 
This logic, however, neglects
possible couplings between the axion and the inflaton sector.
If they do not violate the axion shift symmetry, 
there is no {\it a priori} reason to forbid direct interactions between
the two scalars from the point of view of an effective field theory.

The cosmological consequences of an inflaton-axion coupling was
explored in Ref.~\cite{Kobayashi:2019eyg}, where we demonstrated that
a dimension-four kinetic mixing can lead to a production of axion dark
matter even with a low scale inflation of
(\ref{eq:1.1}).\footnote{For studies of kinetic mixings among multiple
axions, see e.g.
\cite{Babu:1994id,Cicoli:2012sz,Higaki:2014qua,Cicoli:2017zbx,Agrawal:2017cmd}.}
The mechanism proceeds as follows.
The axion during inflation is stabilized close to the bottom of its
potential, but is kicked out at the end of inflation due to the kinetic 
coupling as the inflaton rapidly rolls towards its vacuum. 
This process displaces the axion field from the vacuum and 
sources axion dark matter. 
The inflaton-axion system can also be studied in the field basis where
the kinetic and mass terms are diagonalized.
If the inflaton mass at the vacuum is
larger than the axion mass, then a consistent post-inflation
cosmological history can be realized where
the heavier of the two diagonal fields decays and reheats the universe,
while the lighter one survives and serves as dark matter.
The reheaton and dark matter fields are both linear combinations
of the inflaton and axion, hence are dubbed the {\it inflaxions}.

In the study of the inflaxion mechanism in Ref.~\cite{Kobayashi:2019eyg}, 
we primarily focused on cases where the axion potential stays constant
during the cosmic evolution. However the potential can also vary in time
if it arises from a coupling with a strong gauge group, as is the case
for the QCD axion. 
In this work we explore the possibility that 
after low scale inflation,
the cosmic temperature during the reheating phase 
exceeds the confinement scale~$\Lambda$, and hence the
axion potential temporarily vanishes.
This gives rise to rich dynamics of the inflaton-axion system, 
allowing for a new cosmological scenario for axion dark matter.
Here, the reheating phase plays a central role in the 
dark matter production, 
and we show how this opens up new regions of the axion parameter space. 
We study the implications for both the QCD axion, and
axion-like particles coupled to a hidden confining gauge sector.

This paper is organized as follows:
We start by reviewing the conventional vacuum misalignment scenario in
Sec.~\ref{sec:vm}.
Then we discuss the inflaxion mechanism in Sec.~\ref{sec:inflaxion},
followed by a study of its parameter space in Sec.~\ref{sec:PM}.
We then conclude in Sec.~\ref{sec:conc}.
Technical calculations are relegated to the appendices:
The onset of the axion oscillation is analyzed in detail in
Appendix~\ref{app:onset}. 
The full expressions for the diagonal basis of the inflaxion Lagrangian
are listed in Appendix~\ref{app:diagonal}.

\section{Vacuum Misalignment Scenario}
\label{sec:vm}

Let us start by reviewing the conventional vacuum misalignment scenario. 
Throughout this work we denote the axion by $\sigma$,
and consider it to be coupled to some gauge force that becomes
strong in the IR at an energy scale~$\Lambda$. We assume the axion mass
to depend on the cosmic temperature~$T$ as
\begin{numcases}{m_\sigma (T) \simeq }
 \lambda \,  m_{\sigma 0} \left( \frac{\Lambda }{T} \right)^p
       & for $T \gg \Lambda$, \label{TggL} \\
m_{\sigma 0}
       & for $T \ll \Lambda$, \label{TllL}
\end{numcases}
with the zero-temperature mass written as
\begin{equation}
 m_{\sigma 0} =  \xi \, \frac{\Lambda^2}{f}.
\end{equation}
Here $\xi$ is a dimensionless parameter, and $f$ is an axion decay
constant which sets the periodicity of the axion potential as 
$\sigma \cong \sigma + 2 \pi f$. 
For the QCD axion, the parameters take the values
$\Lambda \approx 200\, \mathrm{MeV}$,
$p \approx 4$,
$\lambda \approx 0.1$,
$\xi \approx 0.1$,
and $f$ is the only free parameter. 
However, in order to keep the discussion general, 
we take all the parameters as arbitrary positive numbers.

The vacuum misalignment scenario can work if the inflationary Hubble scale 
lies within the range 
\begin{equation}
 m_\sigma (T_{\mathrm{inf}}) < H_{\mathrm{inf}} < 2 \pi f,
\label{Hinf-vm}
\end{equation}
where $T_{\mathrm{inf}} = H_{\mathrm{inf}} / 2\pi $ is the de Sitter
temperature during inflation.
The upper bound indicates that the U(1)~symmetry is already
broken\footnote{The symmetry breaking scale can be different from the
scale of the axion periodicity, however we suppose the two scales to be
of the same order throughout this paper.} 
in the inflation epoch and thus the axion field becomes homogeneous in
the observable patch of the universe.
The cosmic temperature during reheating should also satisfy the same
upper bound, $T < f$, to ensure that the symmetry stays broken in the
post-inflation universe. 
The lower bound on the inflation scale indicates that any initial field 
displacement~$\sigma_\star$ of the axion from its potential
minimum stays frozen during inflation due to the Hubble friction.
The axion continues to stay frozen in the post-inflationary universe
while $m_\sigma  < H$, 
but eventually starts to oscillate about its potential minimum when 
the Hubble scale becomes as small as $m_\sigma > H$.
The oscillating field corresponds to a collection of axion particles with a
high occupation number and very small momentum.
The particle number is then approximately conserved,
and the physical number density can be written as
\begin{equation}
 n_\sigma \simeq \frac{1}{2} m_\sigma (T_{\mathrm{osc}})\, 
\sigma_\star^2 \left( \frac{a_{\mathrm{osc}}}{a} \right)^3
\qquad \mathrm{for} \quad
m_\sigma \gg H,
\label{n_sigma}
\end{equation}
in terms of quantities at the onset of the field oscillation
which are denoted by the subscript~$(\mathrm{osc})$. 
Given that the axion starts to oscillate during radiation
domination\footnote{For discussions on cases where radiation domination
takes over after the onset of the oscillations, see 
\cite{Lazarides:1987zf,Banks:1996ea,Giudice:2000ex}.} at
temperatures $T \gg \Lambda$ (i.e. when $m_\sigma \propto T^{-p}$),
then the `onset' of the oscillation can be defined as when the ratio
between the axion mass and the Hubble scale becomes
\begin{equation}
\frac{m_\sigma (T_{\mathrm{osc}})}{H_{\mathrm{osc}}}  
=   (2p+4)
\left[
\pi^{-\frac{1}{2}}\, 
\Gamma \left(\frac{2 p+5}{2 p+4} \right)
\right]^{\frac{2 p+4}{p+3}}
\equiv c_p.
\label{c_pRD}
\end{equation}
This definition of the onset renders the expression~(\ref{n_sigma}) for
the number density exact in the asymptotic future $ a \to \infty$, 
as shown in Appendix~\ref{app:onset}
($c_p$ corresponds to (\ref{eq:mH_osc}) with the substitution~$w = 1/3$).
For instance, the QCD axion with $ p \approx 4$ exhibits $c_4 \approx 4$. 
The Hubble scale during radiation domination is related to the
cosmic temperature via
$ 3 M_{\mathrm{Pl}}^2 H^2 \simeq \rho_\mathrm{r} = (\pi^2/30) g_*(T) T^4 $,
which can be combined with 
(\ref{TggL}) and (\ref{c_pRD}) to give the temperature at the onset of
the oscillation as\footnote{Throughout this paper $M_{\mathrm{Pl}}$
refers to the reduced Planck mass $(8 \pi G)^{-1/2}$.}
\begin{equation}
 T_{\mathrm{osc}} \simeq \Lambda 
\left\{  \left(\frac{\pi^2}{90} g_* (T_\mathrm{osc})
		  \right)^{-1/2} 
\frac{\lambda \xi}{c_p} \frac{M_{\mathrm{Pl}}}{f}
\right\}^{\frac{1}{p+2}}.
\label{T_osc}
\end{equation}

\begin{figure}[t]
\begin{center}
 \begin{minipage}{.45\linewidth}
  \begin{center}
 \includegraphics[width=\linewidth]{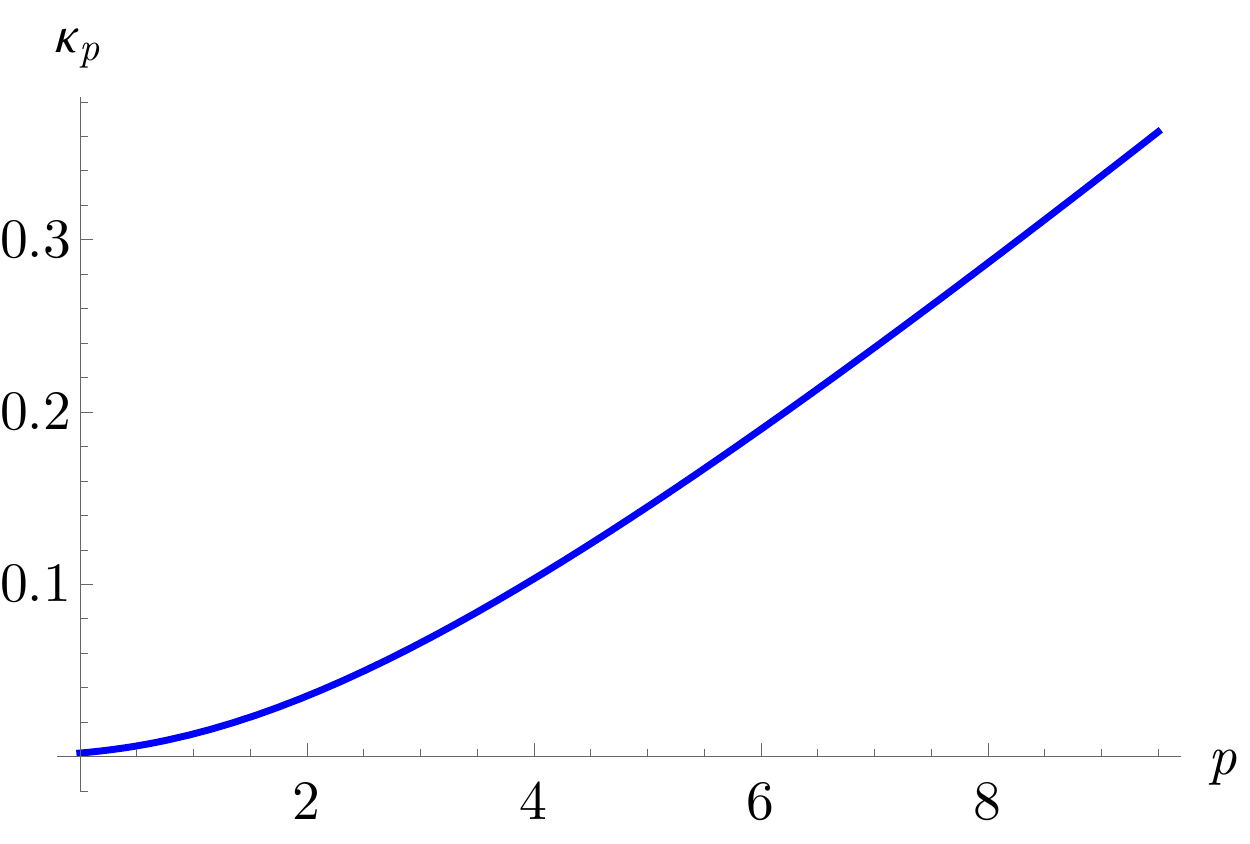}
  \end{center}
 \end{minipage} 
 \begin{minipage}{0.04\linewidth} 
  \begin{center}
  \end{center}
 \end{minipage} 
 \begin{minipage}{.44\linewidth}
  \begin{center}
 \includegraphics[width=\linewidth]{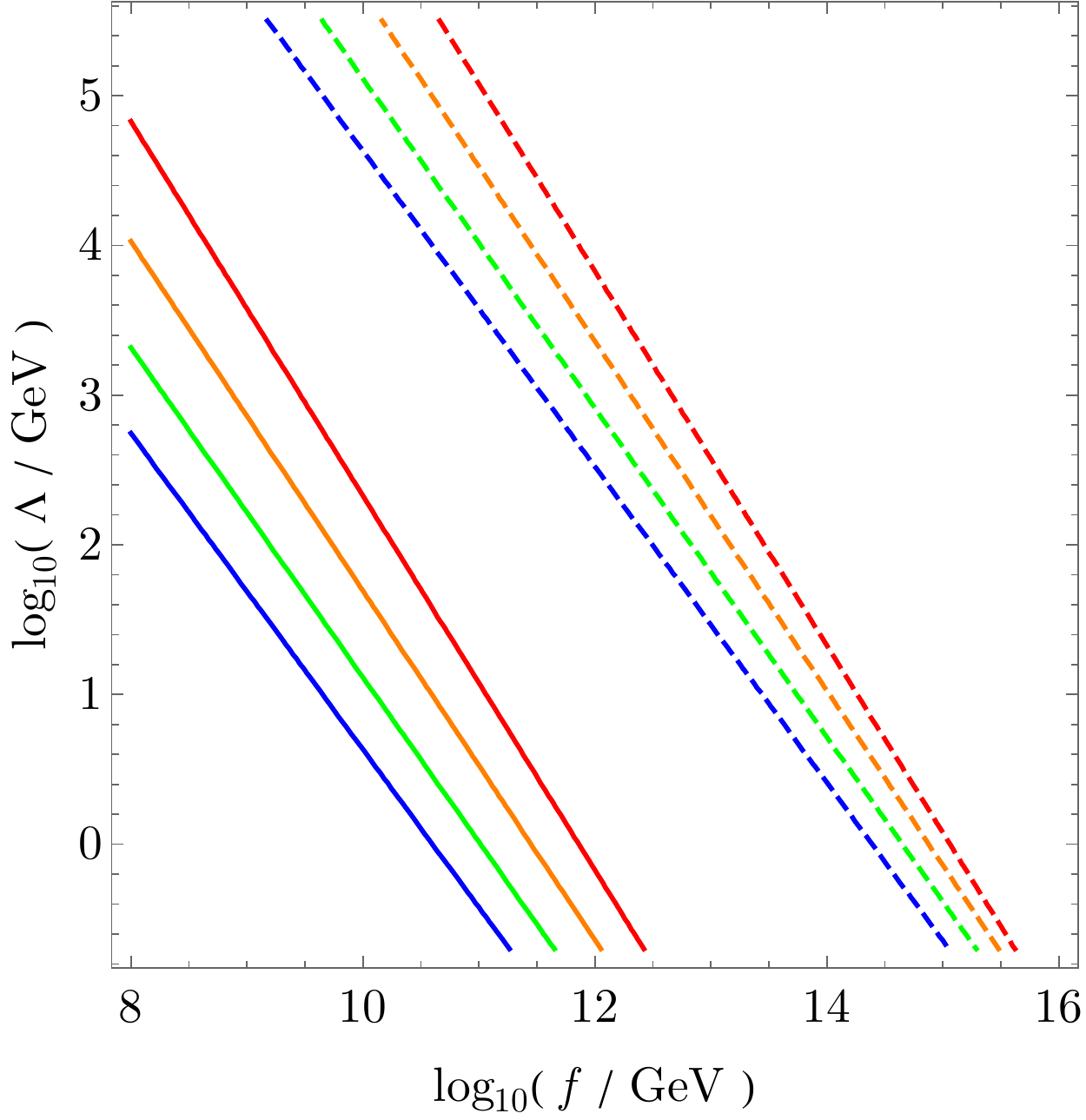}
  \end{center}
 \end{minipage} 
 \caption{Left: Value of the numerical factor~$\kappa_p$ for the axion
 abundance~(\ref{omega-vm}), as a function of the power~$p$ of the
 temperature dependence.
Right: Contours in the plane of the axion decay constant~$f$ and strong
 coupling scale~$\Lambda$ that give rise to the observed dark matter
 abundance  from a vacuum misalignment, for values of the power 
 $p = $ $2$ (red), $4$ (orange), $8$ (green), $16$ (blue), 
 and misalignment angle $\theta_\star = 1$ (solid lines), $10^{-2}$
 (dashed lines).
 Other parameters are fixed to $\lambda = 0.1$, $\xi = 0.1$, 
 $g_* (T_{\mathrm{osc}}) = g_{s*}(T_{\mathrm{osc}}) = 100$. The lower
 edge of the plot represents $\Lambda = 200 \mathrm{MeV}$.}
 \label{fig:kappa_p}
\end{center}
\end{figure}

Considering the entropy of the universe to be conserved since the onset of
the oscillation, the entropy density
$ s = (2 \pi^2/45) g_{s*} (T) T^3 \propto a^{-3}$
can be used to express the axion's number density in the current
universe as 
\begin{equation}
 n_{\sigma 0}  = 
\frac{1}{2} m_\sigma (T_{\mathrm{osc}}) \sigma_\star^2 
\frac{s_0}{s_{\mathrm{osc}}},
\end{equation}
where the subscript~$0$ represents quantities today.
Supposing $T_0 \ll \Lambda$, the present-day axion density is
$\rho_{\sigma 0 } = m_{\sigma 0 } n_{\sigma 0 } $,
and thus by combining with the equations above
one can obtain the density parameter as
\begin{multline}
\qquad
 \Omega_\sigma h^2
= \kappa_p \, \theta_\star^2
\left( \frac{g_{s*}(T_\mathrm{osc})}{100} \right)^{-1}
\left( \frac{g_{*}(T_\mathrm{osc})}{100}\right)^{\frac{p+3}{2p+4}}
\left( \frac{\lambda }{0.1} \right)^{-\frac{1}{p+2}} 
\left( \frac{\xi }{0.1} \right)^{\frac{p+1}{p+2}} 
\\ \times
\left( \frac{\Lambda }{200\, \mathrm{MeV}}\right)
\left(\frac{f}{10^{12}\, \mathrm{GeV}}\right)^{\frac{p+3}{p+2}}.
\qquad
\label{omega-vm}
\end{multline}
Here $\theta_\star \equiv \sigma_\star / f$ is the initial misalignment angle. 
$\kappa_p$ is a numerical factor that depends on the power~$p$,
whose value is plotted in the left panel of Figure~\ref{fig:kappa_p};
for instance, $\kappa_4 \approx 0.1$.
Combinations of $f$ and $\Lambda$ that  yield the observed dark matter
abundance, $\Omega_\sigma h^2 \approx 0.1$~\cite{Aghanim:2018eyx},
are shown in the right panel of Figure~\ref{fig:kappa_p}.

By taking the $p \to 0$ limit (which entails $\kappa_{p \to 0} \approx
0.002$) along with $\lambda \to 1$,
(\ref{omega-vm}) reduces to the relic abundance of an axion with a
constant mass~$m_{\sigma 0}$
(cf., e.g., Eq.~(3.10) of~\cite{Kobayashi:2017jcf}). 
On the other hand, as $\kappa_p$ is a monotonically increasing function
of~$p$, the abundance (\ref{omega-vm}) is enhanced for a large~$p$.
However it should also be noted that upon deriving this result, the
axion was assumed to start oscillating while its mass varies with the
temperature as $m_\sigma \propto T^{-p}$. 
This amounts to assuming that the axion mass at the onset of the
oscillation is smaller than the zero-temperature mass, 
i.e. $\lambda (\Lambda / T_{\mathrm{osc}})^p < 1$,
which combined with (\ref{T_osc}) translates into an upper bound on the
decay constant,
\begin{equation}
 f < \left( \frac{\pi^2}{90}g_{*} (T_{\mathrm{osc}}) \right)^{-1/2}
c_p^{-1} \lambda^{-\frac{2}{p}} \xi M_{\mathrm{Pl}}.
\label{iiid}
\end{equation}
This condition is satisfied for all values of $f$ plotted 
in the right panel of Figure~\ref{fig:kappa_p}.
One can also check that the condition breaks down at large values of~$p$,
and hence, of course, the relic abundance actually does not increase
indefinitely with~$p$.

We should also remark that we have ignored self-interactions of the
axion. Since the axion potential is periodic as
$\sigma \cong \sigma + 2 \pi f$, the relic abundance would receive
anharmonic corrections when the misalignment angle is as large as
$\abs{\theta_\star} \gtrsim 1$ \cite{Turner:1985si,Bae:2008ue}.

Finally, we note that when the axion makes up a significant fraction of
the dark matter in our universe, 
the upper limit of the inflation scale window (\ref{Hinf-vm}) for the vacuum
misalignment scenario becomes much more restrictive due to 
observational constraints on dark matter isocurvature perturbations 
(see e.g. \cite{Kobayashi:2013nva}).

\section{Inflaxion Scenario}
\label{sec:inflaxion}

\subsection{Basic Setup}

It was discovered in \cite{Kobayashi:2019eyg} that even when the
inflation scale is as low as
\begin{equation}
 H_{\mathrm{inf}} < m_{\sigma} (T_{\mathrm{inf}}),
\end{equation}
axion dark matter can be produced by invoking a kinetic mixing between
the axion and the inflaton.\footnote{Variants of this mechanism can also
be constructed with a potential coupling that respects the axion's
discrete shift symmetry, or a coupling of the axion to other fields  
such as the waterfall field in hybrid inflation~\cite{Linde:1991km}.}
The basic idea is captured by the following theory:
\begin{equation}
  \frac{\mathcal{L}}{\sqrt{-g} } =
-\frac{1}{2} g^{\mu \nu} \partial_\mu \sigma  \partial_\nu \sigma 
- \frac{1}{2} m_\sigma(T)^2 \sigma^2 
-\frac{1}{2} g^{\mu \nu} \partial_\mu \phi  \partial_\nu \phi 
- V(\phi) 
- \alpha \, g^{\mu \nu} \partial_\mu \phi \partial_\nu \sigma 
+ L_{\mathrm{c}}[\sigma, \phi, \Psi].
\label{eq:Lagrangian}
\end{equation}
Here, $\sigma$ is the axion whose mass term is understood to arise from
expanding the periodic potential around one of the minima,  
$\phi$ is the inflaton with a potential~$V(\phi)$ that possesses an
inflationary plateau, 
$\alpha$ is a dimensionless coupling constant that satisfies
$\abs{\alpha} < 1$ to avoid ghost degrees of freedom,
and $L_{\mathrm{c}}$ represents couplings with other matter fields 
which we collectively denote by~$\Psi$. 
Given that the axion is a pseudoscalar, the inflaton would also need to be a
pseudoscalar for the kinetic mixing term to conserve parity;
however we remark that parity conservation is not a
prerequisite for the mechanism to operate.

The main part of the analysis in \cite{Kobayashi:2019eyg} was
devoted to axions with a constant mass;
for axions coupled to a strong sector,
this amounts to assuming that the cosmic temperature never exceeds
the strong coupling scale~$\Lambda$.
In the following, we instead analyze the case where the temperature
in the post-inflation universe becomes higher than~$\Lambda$, but lower
than~$f$, so that the axion mass temporarily diminishes, while the
U(1)~symmetry continues to be broken.
To be concrete, we consider the axion mass to depend on the temperature
as (\ref{TggL}) and (\ref{TllL}),
and focus on cases where the inflationary de Sitter
temperature~$T_{\mathrm{inf}}$ and the maximum temperature 
of radiation $T_{\mathrm{max}}$ 
during the reheating process satisfy 
\begin{equation}
 T_{\mathrm{inf}} < \Lambda < T_{\mathrm{max}} < f.
\label{ro4}
\end{equation}
As in \cite{Kobayashi:2019eyg}, the inflationary Hubble scale is
considered to be smaller than the zero-temperature axion mass, which in
turn is smaller than the inflaton mass~$m_{\phi 0}$ at the vacuum.
In the following we further assume that the axion mass becomes smaller
than the Hubble scale when the temperature
reaches~$T_{\mathrm{max}}$.\footnote{We have in mind here 
perturbative reheating in which $T_{\mathrm{max}}$ is reached
within about a Hubble time after the end of inflation. 
Hence the Hubble rate upon $T = T_{\mathrm{max}}$ 
is of the same order as the Hubble rate at the end of inflation.}
Thus we impose the following hierarchy:
\begin{equation}
 m_\sigma (T_{\mathrm{max}}) < H_{\mathrm{inf}}
 < m_{\sigma 0} < m_{\phi 0}.
\label{hierarchy}
\end{equation}

\begin{figure}[t]
\begin{center}
 \begin{minipage}{.47\linewidth}
  \begin{center}
 \includegraphics[width=\linewidth]{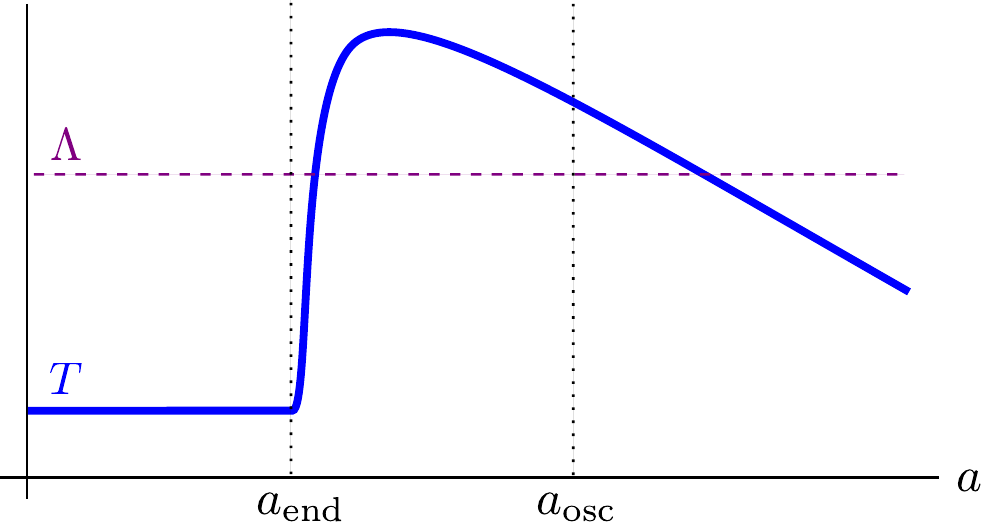}
  \end{center}
 \end{minipage} 
 \begin{minipage}{0.01\linewidth} 
  \begin{center}
  \end{center}
 \end{minipage} 
 \begin{minipage}{.47\linewidth}
  \begin{center}
 \includegraphics[width=\linewidth]{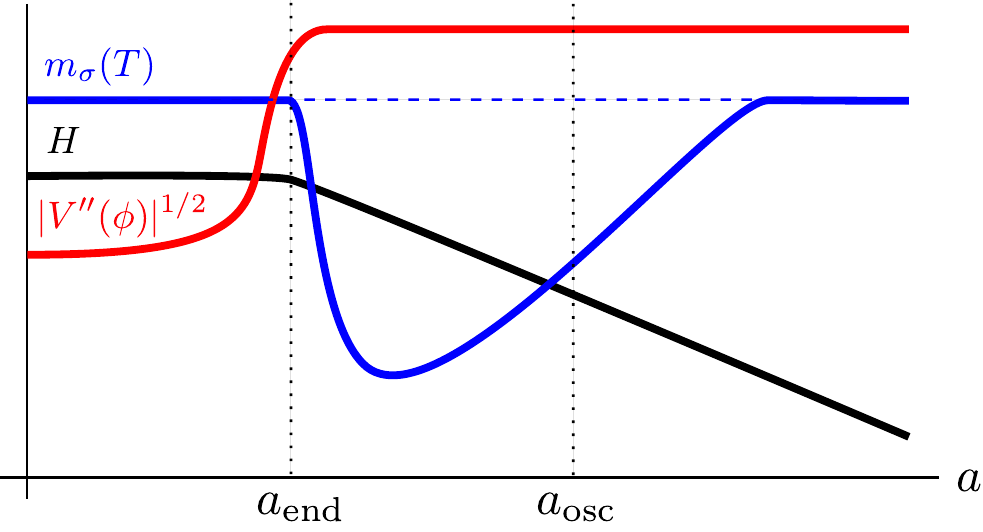}
  \end{center}
 \end{minipage} 
 \caption{Schematic of the time evolution of the cosmic temperature
 (left) and scalar field masses (right) in the inflaxion scenario under
 consideration (not to scale).}
 \label{fig:m-a}
\end{center}
\end{figure}

The time evolution of the temperature and the scalar field masses
are illustrated in Figure~\ref{fig:m-a}. 
Here, $a_{\mathrm{end}}$ represents the scale factor when inflation ends.
In the left panel, the cosmic temperature during inflation is taken as
the de Sitter temperature, while after inflation it is the
radiation temperature arising from the decay of the reheaton.
(The de Sitter and radiation temperatures are connected at the end of
inflation for illustration purposes only.)
The right panel shows the evolution of the Hubble rate~$H$ (black curve),
the axion mass~$m_\sigma (T)$ (blue), and the effective mass
$\abs{V''(\phi)}^{1/2}$ of the inflaton (red). 
The axion mass during inflation takes its zero-temperature
value~$m_{\sigma 0}$, then during reheating becomes smaller than~$H$ for
a while, and again becomes~$m_{\sigma 0}$ in the later universe when
$T \ll \Lambda$. 
The inflaton potential~$V(\phi)$ is considered to possess a plateau
that enables slow-roll inflation, 
and hence $\abs{V''(\phi)}^{1/2} < H$ during inflation.
This inequality breaks down towards the end of inflation, as the
inflaton rolls to its vacuum.
Inflation thus ends and the inflaton starts to oscillate around the 
minimum of its potential, which we assume to be 
approximated by a quadratic,
\begin{equation}
 V(\phi) \simeq \frac{1}{2} m_{\phi 0}^2 \phi^2
\qquad \mathrm{for} \quad
\abs{\phi} \leq \abs{\phi_{\mathrm{end}}}.
\label{Vquad}
\end{equation}
Here, $\phi_{\mathrm{end}}$
refers to the inflaton field value where inflation ends. 
The inflaton mass thus becomes~$m_{\phi 0}$,
which is larger than~$m_{\sigma 0}$ as required in~(\ref{hierarchy}).
We have in mind here small-field inflation models in which
$\abs{V''(\phi)}^{1/2}$ takes very different values between the
plateau region and the minimum. 
However we should also note that this transition of $\abs{V''(\phi)}^{1/2}$
from a tiny value during inflation to a larger $m_{\phi 0}$ is not
necessarily monotonic as shown in the simplified illustration;
$\abs{V''(\phi)}^{1/2}$ can instead oscillate due to higher order terms
in the potential while the inflaton's oscillation amplitude is large.

\subsection{End of Inflation and Reheating}

The post-inflationary dynamics of the inflaton and axion is
insensitive to the details of the inflation model,
and thus we start our discussion from the time when inflation ends.

Let us for a moment ignore the temperature dependence of the axion
mass. Then one can simultaneously diagonalize the kinetic terms as well
as the mass terms so that the Lagrangian~(\ref{eq:Lagrangian}) with the
quadratic inflaton potential (\ref{Vquad}) is rewritten as
\begin{equation}
  \frac{\mathcal{L}}{\sqrt{-g} } = 
\sum_{i=\mathrm{DM}, \mathrm{RH}} \left(
-\frac{1}{2} g^{\mu \nu} \partial_\mu \varphi_i \, \partial_\nu \varphi_i
- \frac{1}{2} m_i^2 \varphi_i^2 
\right) + L_{\mathrm{c}}[\sigma, \phi, \Psi],
\end{equation}
where the explicit forms of the diagonalized fields and their masses are
given in Appendix~\ref{app:diagonal}.
Here, we suppose a mass hierarchy $m_\sigma^2 \ll m_{\phi 0}^2 $ and
use approximate expressions of 
\begin{equation}\label{eq:diagonal}
\begin{split}
\varphi_{\mathrm{DM}} \simeq
\alpha \, \phi + \sigma,
& \quad
\varphi_{\mathrm{RH}} \simeq
\sqrt{1 - \alpha^2}
\left( \phi  - \alpha  \frac{m_\sigma^2}{m_{\phi 0}^2} \, \sigma \right),
\\
 m_{\mathrm{DM}} \simeq  m_\sigma,
& \quad
 m_{\mathrm{RH}} \simeq \frac{ m_{\phi 0}}{\sqrt{1-\alpha^2}},
\end{split}
\end{equation}
where each of the coefficients of $\phi$ and $\sigma$,
as well as the diagonalized masses are given 
to leading order in a $m_\sigma^2 / m_{\phi 0}^2$ expansion.
($\varphi_{\mathrm{DM}}$ and $\varphi_{\mathrm{RH}}$ correspond
respectively to $\varphi_+$ and $\varphi_-$ in (\ref{varphi_pm})
up to overall signs.)
Recalling $\alpha^2 < 1$, one sees that 
$m_{\mathrm{DM}} \simeq m_\sigma < m_{\phi 0} < m_{\mathrm{RH}}$.
The lighter field $\varphi_{\mathrm{DM}}$ can be long-lived for a
sufficiently small axion mass, and thus serves as a dark matter candidate.
The heavier field~$\varphi_{\mathrm{RH}}$, on the other hand,
can reheat the universe through its decay.

The diagonal basis is also convenient for analyzing the decay of the
scalar particles.
If, for instance, the axion and inflaton were coupled to (either
Standard Model (SM) or hidden)
photons and/or a light Dirac fermion~$\psi$ via 
\begin{equation}
 L_{\mathrm{c}}[\sigma, \phi, \Psi] = 
\frac{G_{\sigma \gamma \gamma }}{4}
\sigma F_{\mu \nu} \tilde{F}^{\mu \nu} 
 +  \frac{G_{\phi \gamma \gamma }}{4}
\phi F_{\mu \nu} \tilde{F}^{\mu \nu} 
+ g_{\phi ff} \, \phi \bar{\psi} i \gamma^5 \psi,
\label{eq:couplings}
\end{equation}
then the decay widths of the dark matter and reheaton are given to
leading order in $m_\sigma^2 / m_{\phi 0}^2$ as
(see also (\ref{eq:pm-to-gg}) and (\ref{eq:pm-to-ff})
for the full expressions)
\begin{gather}
 \Gamma (\varphi_\mathrm{DM} \to \gamma \gamma)
\simeq \frac{G_{\sigma \gamma \gamma }^2 }{64 \pi } m_\sigma^3,
\quad
 \Gamma (\varphi_\mathrm{RH} \to \gamma \gamma )
\simeq 
\frac{\alpha^2}{(1-\alpha^2)^{5/2}}
\frac{G_{\sigma \gamma \gamma }^2 }{64 \pi } m_{\phi 0}^3,
\label{eq:3.9}
\\
 \Gamma (\varphi_\mathrm{DM}  \to \gamma \gamma )
\simeq 
\alpha^2 
\frac{G_{\phi \gamma \gamma }^2 }{64 \pi }
\frac{m_\sigma^7}{m_{\phi 0}^4},
\quad
 \Gamma (\varphi_\mathrm{RH} \to \gamma \gamma )
\simeq 
\frac{1}{(1-\alpha^2)^{5/2}}
\frac{G_{\phi \gamma \gamma }^2 }{64 \pi } m_{\phi 0}^3,
\label{eq:3.10}
\\
  \Gamma (\varphi_\mathrm{DM} \to f \bar{f} )
\simeq \alpha^2 \frac{g_{\phi ff}^2}{8 \pi }
\frac{m_\sigma^5}{m_{\phi 0}^4},
\quad
  \Gamma (\varphi_\mathrm{RH} \to f \bar{f} )
\simeq \frac{1}{(1-\alpha^2)^{3/2}} \frac{g_{\phi ff}^2}{8 \pi }
m_{\phi 0}.
\label{eq:3.11}
\end{gather}
The decay widths in each line are induced by each of the terms in
(\ref{eq:couplings}), and here we have ignored the 
contribution to the two-photon decay rates from the 
cross-term~$\propto G_{\sigma \gamma \gamma } G_{\phi \gamma \gamma }$.
These expressions explicitly show that the life time of the reheaton is
suppressed compared to that of the dark matter field by powers of their
mass ratio.

Hereafter we promote the axion mass in the
expressions of~(\ref{eq:diagonal}) to a temperature-dependent
mass~$m_\sigma (T)$, and analyze the post-inflationary dynamics in terms
of the (pseudo)diagonal fields. 
Later on, we compare the results that follow from this analytic
procedure with 
those obtained by solving the full set of equations of motion. 

The inflaton field value at the end of inflation can be estimated 
by noting that a significant fraction of the total energy density of the
universe is still in the inflaton's potential energy~(\ref{Vquad}),
i.e.,
\begin{equation}
 M_{\mathrm{Pl}}^2 H_{\mathrm{end}}^2 \sim m_{\phi 0}^2
  \phi_{\mathrm{end}}^2, 
\label{eq:phi_end}
\end{equation}
where we use the subscript~$(\mathrm{end})$ to represent quantities at
the end of inflation.
It was found in~\cite{Kobayashi:2019eyg} that the axion field value
becomes comparable to $\alpha \phi$ towards the end of inflation,
and hence the dark matter
field value is obtained as
\begin{equation}
 \varphi_{\mathrm{DM} \, \mathrm{end}}^2 = 
\left( \frac{C \alpha M_{\mathrm{Pl}} H_{\mathrm{end}}}{m_{\phi 0}}
\right)^2.
\label{eq:varphi_DMend}
\end{equation}
Here, $C$ is a numerical factor whose exact value depends on the model,
and is typically of $C \sim 10$. 
The dark matter field begins to oscillate with this initial amplitude, 
with a mass equal to the zero-temperature axion mass, 
i.e. $m_{\mathrm{DM \, end}} \simeq m_{\sigma 0}$,
since the reheaton~$\varphi_{\mathrm{RH}}$ still has not started to decay
at this point. The dark matter field's potential energy is
\begin{equation}
 \left( \frac{1}{2} m_{\mathrm{DM}}^2 \varphi_{\mathrm{DM}}^2
 \right)_{\mathrm{end}} 
 \simeq \frac{C^2 \alpha^2}{6}  \frac{m_{\sigma 0}^2}{m_{\phi0}^2} 
 \times 3 M_{\mathrm{Pl}}^2 H_{\mathrm{end}}^2,
\label{eq:rho_DM_end}
\end{equation}
which is suppressed compared to the total energy density of the universe
by the mass ratio $m_{\sigma 0}^2 / m_{\phi 0}^2$.
This indicates that the post-inflation universe is initially dominated
by the reheaton. 

The reheaton undergoes oscillations and decays into hot radiation, which
forces the dark matter mass to diminish. 
We now evaluate the reheating process without specifying the explicit
forms of the matter couplings. Here we only assume that reheating
proceeds by a perturbative decay of the reheaton
into radiation with a decay width~$\Gamma_{\mathrm{RH}}$. 

If $\Gamma_{\mathrm{RH}} < H_{\mathrm{end}}$, 
then the radiation density would reach its maximum value
\begin{equation}
 \rho_{\mathrm{r\, max}} \sim 
\frac{\Gamma_{\mathrm{RH}} }{H_{\mathrm{end}}}
M_{\mathrm{Pl}}^2 H_{\mathrm{end}}^2,
\end{equation}
within about a Hubble time after the end of
inflation~\cite{Kolb:1990vq}.\footnote{This can be checked
explicitly by solving  the continuity equation for the radiation density,
\begin{equation}
 \dot{\rho_{\mathrm{r}}} + 4 H \rho_{\mathrm{r}} = \Gamma_{\mathrm{RH}}
  \rho_{\mathrm{RH}},
\label{eq:rad-cont}
\end{equation}
with an initial condition $\rho_{\mathrm{r\, end}} = 0$.
Here, the energy density of the decaying reheaton can be written as
\begin{equation}
 \rho_{\mathrm{RH}} = \left(\frac{a}{a_{\mathrm{end}}} \right)^{-3}
e^{-\Gamma_{\mathrm{RH}} (t-t_{\mathrm{end}})} \rho_{\mathrm{RH\, end}},
\end{equation}
and considering the post-inflation universe to be initially dominated by
the non-relativistic reheaton particles
($\rho_{\mathrm{RH\, end}} \simeq 3 M_{\mathrm{Pl}}^2 H_{\mathrm{end}}^2$)
gives a scaling $H^2 \propto a^{-3}$.
Then the solution of the continuity equation, to linear order
in~$\Gamma_{\mathrm{RH}}$, is
\begin{equation}
  \rho_{\mathrm{r}} \simeq \frac{6}{5} 
\frac{\Gamma_{\mathrm{RH}} }{H_{\mathrm{end}}}
M_{\mathrm{Pl}}^2 H_{\mathrm{end}}^2
\left\{ 
\left( \frac{a}{a_{\mathrm{end}}} \right)^{-3/2}
- \left( \frac{a}{a_{\mathrm{end}}} \right)^{-4}
\right\}
\qquad \mathrm{for} \quad 
\Gamma_{\mathrm{RH}} \ll H.
\label{eq:rad3over2}
\end{equation}
At $a_{\mathrm{max}} \approx 1.5 \times a_{\mathrm{end}}$,
this expression takes its maximum value
\begin{equation}
 \rho_{\mathrm{r \, max}} \approx 0.4 \times 
\frac{\Gamma_{\mathrm{RH}}}{H_{\mathrm{end}}}
M_{\mathrm{Pl}}^2 H_{\mathrm{end}}^2 .
\end{equation}
}
Subsequently the radiation density turns to redshift, albeit slowly as
it continues to be sourced by the decaying reheaton, 
and eventually dominates over the reheaton density
when $H \sim \Gamma_{\mathrm{RH}}$. 

If on the other hand $\Gamma_{\mathrm{RH}} > H_{\mathrm{end}}$, 
then the reheaton would quickly decay\footnote{$\Gamma_\mathrm{RH} >
H_{\mathrm{end}}$ after inflation does
not mean that the inflaton fluctuations should have decayed during
inflation, since the effective mass of the inflaton during slow-roll
inflation is much smaller than $m_{\phi 0}$.} 
and radiation domination would
take over right after the end of inflation, yielding
\begin{equation}
 \rho_{\mathrm{r\, max}} \sim M_{\mathrm{Pl}}^2 H_{\mathrm{end}}^2.
\end{equation}

Thus for all cases 
$\Gamma_{\mathrm{RH}} \gtreqless H_{\mathrm{end}}$,
the Hubble rate when radiation domination takes over can be
collectively written as
\begin{equation}
 H_{\mathrm{dom}} \sim \mathrm{min.} \left\{ H_{\mathrm{end}},
     \Gamma_{\mathrm{RH}}  \right\} ,
\label{H_dom}
\end{equation}
and the maximum radiation density as
\begin{equation}
 \rho_{\mathrm{r\, max}} \sim \frac{H_{\mathrm{dom}}}{H_{\mathrm{end}}}
M_{\mathrm{Pl}}^2 H_{\mathrm{end}}^2.
\end{equation}
Therefore the maximum temperature during reheating is written as
\begin{equation}
 T_{\mathrm{max}} \sim 10^4 \, \mathrm{GeV}
\left( \frac{g_* (T_{\mathrm{max}})}{100} \right)^{-1/4}
\left( \frac{H_{\mathrm{end}}}{1\, \mathrm{eV}}\right)^{1/2}
\left( \frac{H_{\mathrm{dom}}}{H_{\mathrm{end}}} \right)^{1/4}.
\label{T_max}
\end{equation}

\subsection{Drifting Away from the Vacuum}
\label{sec:kick}

As the radiation temperature increases after the end of inflation, the
axion mass becomes smaller than the Hubble scale. 
Then the dark matter
field, being effectively massless, streams freely with the velocity
that it had acquired before its mass diminished. 
In this way the field obtains a further displacement from its potential
minimum.\footnote{If instead the radiation
temperature never exceeds~$\Lambda$ and the axion mass stays constant,
then the field displacement (\ref{eq:varphi_DMend}) at
the very end of inflation would be the only source for dark matter
production~\cite{Kobayashi:2019eyg}.}

To make a rough estimate of this effect, note that at the
end of inflation when the radiation temperature is effectively
zero,\footnote{Here perturbative reheating after inflation is assumed.
However the radiation temperature may rise already before the inflaton
begins to oscillate, if, for instance, tachyonic
preheating~\cite{Felder:2000hj} takes place. It would be
interesting to explore the inflaxion mechanism in such cases as well.} 
the dark matter field is beginning to oscillate with 
the zero-temperature axion mass
$m_{\mathrm{DM\, end}} \simeq m_{\sigma 0}$.
The field velocity at this time is thus estimated as
\begin{equation}
 \abs{\dot{\varphi}_{\mathrm{DM\, end}} }
\sim m_{\sigma 0} \abs{\varphi_{\mathrm{DM\, end}}},
\label{eq:ini-vel}
\end{equation}
where an overdot denotes a derivative with respect to physical time.
Then if the temperature rises rapidly and hence
the axion mass, or equivalently the dark matter mass,
immediately vanishes, the dark matter field would 
begin to free-stream with the above initial velocity.
However the Hubble friction damps the velocity of a free field, and
so the dark matter field comes to a halt after a few Hubble
times.\footnote{The velocity of a massless homogeneous field redshifts as
$\dot{\varphi}_{\mathrm{DM}} \propto a^{-3}$. Integrating this from the
end of inflation in a 
universe with a constant equation of state~$w$ $(\neq 1)$ yields
\begin{equation}
 \varphi_{\mathrm{DM}} - \varphi_{\mathrm{DM\, end}} 
= \frac{2}{3 (1-w)} \frac{\dot{\varphi}_{\mathrm{DM\,
end}}}{H_{\mathrm{end}}} 
\left\{  1 - \left( \frac{a}{a_{\mathrm{end}}} \right)^{-\frac{3
(1-w)}{2}} \right\} .
\label{eq:exact-fs}
\end{equation}
} 
Hence the field moves a distance of 
\begin{equation}
 \abs{\Delta \varphi_{\mathrm{DM}}} \sim 
\left| \frac{\dot{\varphi}_{\mathrm{DM\, end}}}{H_{\mathrm{end}}} \right|
\sim \frac{m_{\sigma 0}}{H_{\mathrm{end}}} 
\abs{\varphi_{\mathrm{DM\, end}}}
> \abs{\varphi_{\mathrm{DM\, end}}},
\label{eq:free-stream}
\end{equation}
where we used (\ref{eq:ini-vel}) in the second approximation, and 
(\ref{hierarchy}) for the last inequality.
This indicates that the field excursion during the free-streaming
dominates over the field displacement at the end of inflation.
Hence by combining (\ref{eq:free-stream}) with (\ref{eq:varphi_DMend}), 
the dark matter field value after the free-streaming is obtained
as 
\begin{equation}
 \varphi_{\mathrm{DM} \star}^2 = 
\left(
\frac{ B \alpha M_{\mathrm{Pl}} m_{\sigma 0}}{m_{\phi 0}}
\right)^2.
\label{eq:fin-dis}
\end{equation}
Here, $B$ is a dimensionless parameter which is of order unity 
according to the discussions above.\footnote{(\ref{eq:ini-vel}) would
overestimate the initial velocity if $\varphi_{\mathrm{DM}}$ at the end
of inflation is just about to start oscillating, and
(\ref{eq:free-stream}) can also overestimate the free-streaming
distance, as one sees by comparing with the exact
expression~(\ref{eq:exact-fs}). Considering these to be
compensated by the factor $C \sim 10$ in (\ref{eq:varphi_DMend}) yields
a naive estimate of $B \sim 1$.}
However we should also note that
this is only a crude approximation and the actual field
dynamics can be more intricate.
For instance, if $m_{\sigma 0}$ is only marginally larger
than~$H_{\mathrm{end}}$, then the free-streaming
distance~$\Delta \varphi_{\mathrm{DM}}$ and the initial
displacement~$\varphi_{\mathrm{DM\, end}}$ would be comparable in size
and thus might cancel each other, yielding a much smaller
field displacement.
The approximation that the field begins to free-stream with the
initial velocity~(\ref{eq:ini-vel}) could also break down, 
if the time scale~$\Delta t$ for the axion mass to diminish
is larger than the initial oscillation period, 
i.e. $\Delta t \gtrsim 2 \pi / m_{\sigma 0}$;
cases with a gradually decreasing mass will be discussed in detail in
the next subsection where we numerically study concrete examples.
All such effects that give corrections to the simplest picture discussed
above would amount to shifting the parameter~$B$ from order unity. 

After the free-streaming, the dark matter field stays
frozen at $\varphi_{\mathrm{DM}\star}$, and then begins to oscillate
about its potential minimum as the cosmic temperature decreases and the
axion mass again becomes larger than the Hubble rate. 
The field dynamics hereafter is the same as in the conventional vacuum
misalignment scenario. 
In this sense, the temperature-dependent inflaxion can be 
considered as a mechanism that sources a vacuum misalignment
of~(\ref{eq:fin-dis}) with low scale inflation. 
Hence we can apply the results of Section~\ref{sec:vm} to compute the
dark matter abundance:
Assuming radiation domination to take over before the dark matter field
starts to oscillate, i.e.
\begin{equation}
 H_{\mathrm{dom}} > H_{\mathrm{osc}},
\label{eq:dom-osc}
\end{equation}
and also the axion mass to be varying with temperature at the onset
of the oscillation, i.e. (\ref{iiid}),
then the present-day dark matter abundance can be computed as
(\ref{omega-vm}), with the misalignment angle given by
\begin{equation}
 \theta_{\star}^2 = \left( \frac{\varphi_{\mathrm{DM} \star}}{f}\right)^2
= \left(\frac{B \alpha M_{\mathrm{Pl}} m_{\sigma 0}}{f m_{\phi 0}} \right)^2.
\label{eq:theta_star}
\end{equation}
It is worth stressing that, unlike in the conventional vacuum
misalignment scenario 
where the angle is given as a random initial condition,
here it is uniquely fixed by the inflaxion parameters.
We also note that the angle in this inflaxion scenario
is independent of the inflation scale.

By the time the dark matter field starts to oscillate, the reheaton has
decayed away and thus the field value of the inflaton is much smaller than
that of the axion, i.e. $\abs{\phi} \ll \abs{\sigma}$,
as can be seen by setting $\varphi_{\mathrm{RH}} \approx 0$ in
(\ref{eq:diagonal}). This in turn suggests that the dark matter degree
of freedom becomes dominated by the axion field,
$\varphi_{\mathrm{DM}} \simeq \sigma$.
We also note that we have neglected the axion's self-interactions in our
analyses, which is justified if the final displacement of the
axion dark matter field~$\abs{\varphi_{\mathrm{DM} \star}}$ is
sufficiently smaller 
than the periodicity of the axion potential~$2 \pi f $,
i.e.,
\begin{equation}
 \abs{\theta_\star} \lesssim 1.
\label{eq:harmonic}
\end{equation}
If instead the field displacement exceeds half the periodicity, 
$\abs{\varphi_{\mathrm{DM} \star}} > \pi f$,
then the axion field would get trapped not in the minimum around which
we have been expanding the axion potential, but
in another minimum that lies near~$\varphi_{\mathrm{DM} \star}$. 
The distance to this nearby minimum at the onset of the oscillation 
would typically be $\sim f$, 
hence the relic abundance in such cases is given by 
(\ref{omega-vm}) with a misalignment angle $\abs{\theta_\star} \sim 1$. 
Anharmonic corrections to the abundance computation of~(\ref{omega-vm})
would become important if the axion, after the free-streaming, lands on
a point that happens to be close to a potential maximum.

We should also remark that there is no dark matter isocurvature
perturbation in our inflaxion scenario, 
since $m_{\sigma 0} > H_{\mathrm{inf}}$ and so inflation is
effectively single-field.

\subsection{Numerical Examples}

In this subsection we study the inflaton-axion dynamics in concrete
examples by numerically solving the full set of equations 
of motion in a flat FRW universe.
The coupled equations of motion of the homogeneous inflaton and axions
fields that incorporate the decay of the scalar particles
as effective friction terms are given in (\ref{eq:fullEoMs}) in
Appendix~\ref{app:diagonal}.
The total energy-momentum tensor of the homogeneous fields can be
written in the form of a perfect fluid,
\begin{equation}
 T_{\mu \nu}^{\sigma \phi} = \rho_{\sigma \phi} u_\mu u_\nu
+ p_{\sigma \phi} (g_{\mu \nu} + u_\mu u_\nu),
\end{equation}
where $u^\mu$  is a velocity vector normalized as
$u_\mu u^\mu = -1$, with its spatial components
vanishing in the Cartesian coordinates, $u^i = 0$.
The energy density and pressure of the inflaton-axion system
is
\begin{equation}
\begin{split}
 \rho_{\sigma \phi} &= \frac{1}{2} \dot{\sigma}^2 + 
\frac{1}{2} m_\sigma^2\sigma^2 + \frac{1}{2} \dot{\phi}^2 + V(\phi)
+ \alpha \dot{\phi} \dot{\sigma},
\\
 p_{\sigma \phi} &= \frac{1}{2} \dot{\sigma}^2 -
\frac{1}{2} m_\sigma^2\sigma^2 + \frac{1}{2} \dot{\phi}^2 - V(\phi)
+ \alpha \dot{\phi} \dot{\sigma}.
\end{split}
\end{equation}
In the numerical computation we assume all the other components of
the universe to be thermalized and to form a radiation
fluid.\footnote{Since the axion mass arises from a coupling with 
a gauge force, a derivative of such a term with the gauge field or
the metric also contributes to the total energy-momentum tensor. 
Here we include such contributions, and also those arising 
from~$L_\mathrm{c}$, into the ``radiation component''.} 
Hence the Friedmann equation reads
\begin{equation}
 3 M_{\mathrm{Pl}}^2 H^2 = \rho_{\sigma \phi} + \rho_{\mathrm{r}},
\label{eq:Friedmann}
\end{equation}
and the continuity equation is
\begin{equation}
 \dot{\rho}_{\sigma \phi} + \dot{\rho}_{\mathrm{r}}
+ 3 H \left( \rho_{\sigma \phi} + p_{\sigma \phi} \right)
+ 4 H \rho_{\mathrm{r}}
= 0.
\label{eq:continuity}
\end{equation}

During the inflationary epoch, we ignore particle decay and set the
radiation density to be negligible. Moreover, by considering a low scale
inflation such that $T_{\mathrm{inf}} \ll \Lambda$, the axion mass is
fixed to the zero-temperature value.
Hence we solve the set of equations
(\ref{eq:fullEoMs}) and (\ref{eq:Friedmann}) by setting
$\Gamma_\pm = 0$, $\rho_{\mathrm{r}} = 0$, $m_\sigma = m_{\sigma 0}$. 

Then, when the cosmological expansion turns from an acceleration to a
deceleration, i.e. when $\dot{H} / H^2 = -1$,
we include the decay widths into the fields' equations of
motion~(\ref{eq:fullEoMs}), which are thereafter solved along with 
(\ref{eq:Friedmann}) and (\ref{eq:continuity}).
The temperature dependence of the axion mass is modeled as
$m_\sigma(T) = m_{\sigma 0} \tanh [ \lambda (\Lambda / T)^p ]$
to reproduce the asymptotic behaviors (\ref{TggL}) and (\ref{TllL}),
and $T$ is set to the radiation temperature determined via
$ \rho_\mathrm{r} = (\pi^2/30) g_* T^4 $.

As a toy inflaton potential that smoothly connects between an
inflationary plateau with the minimum~(\ref{Vquad}), we studied a
potential of the form
\begin{equation}
 V(\phi) = m_{\phi 0}^2 \mu^2 
\left( 1 - \frac{2}{e^{\phi/\mu} + e^{-\phi/\mu}}  \right).
\label{eq:sech}
\end{equation}
The inflation scale for this potential is 
$H_{\mathrm{inf}} \simeq m_{\phi 0} \mu / \sqrt{3} M_{\mathrm{Pl}}$
until the end of inflation,
and we adopted the value $\mu = 4 \times 10^{14}\, \mathrm{GeV}$ so that 
$m_{\phi 0 }  \approx 10^4 H_{\mathrm{end}} $.
The axion mass was chosen as
$m_{\sigma 0} = 10^{-2} m_{\phi 0 } \approx 10^2 H_{\mathrm{end}} $
with 
$\lambda = 10^{-1}$,
$p = 6$, 
$\Lambda = 10^{-1} \cdot  (3 M_{\mathrm{Pl}}^2 H_{\mathrm{end}}^2)^{1/4} $.
The kinetic coupling was taken to be nonzero, but much smaller
than unity, $0 < \abs{\alpha} \ll 1$.
We also fixed the relativistic degrees of freedom to a constant value
$g_* = 50$ for simplicity.
These parameters were chosen mainly for the purpose of reducing
the computational time.
The reader will have noticed that the exact values are not
specified for $\alpha$, $m_{\phi 0}$, $m_{\sigma 0}$, etc.; 
this is because the plots we show below in terms of dimensionless
quantities are independent of the exact values of such parameters.
Regarding the decay channel,
we considered a coupling between the inflaton and fermions as
$ L_{\mathrm{c}} =  g_{\phi ff} \, \phi \bar{\psi} i \gamma^5 \psi $,
and used the expressions~(\ref{eq:pm-to-ff}) for the decay widths.
We have performed the computation for several
different values of the dimensionless coupling $g_{\phi ff}$.

\begin{figure}[t]
\begin{center}
 \begin{minipage}{.46\linewidth}
  \begin{center}
 \includegraphics[width=\linewidth]{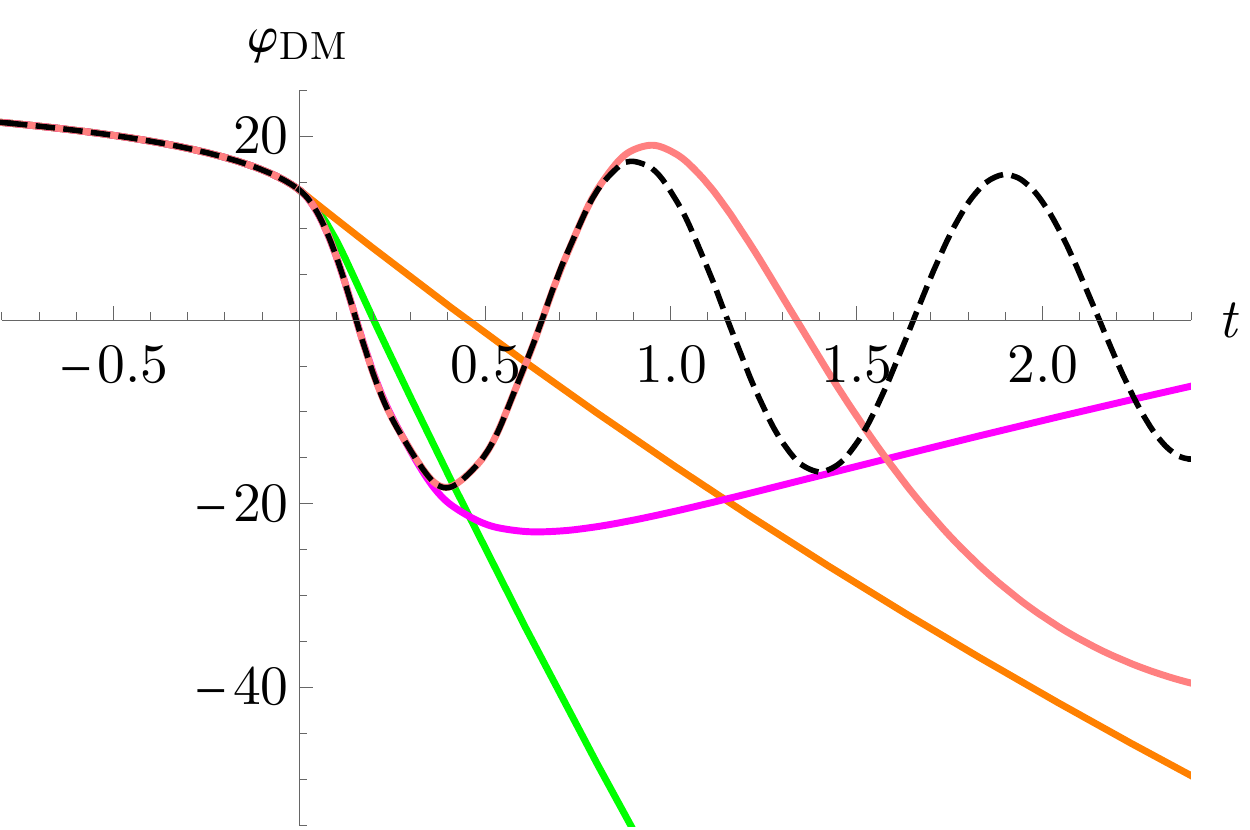}
  \end{center}
 \end{minipage} 
 \begin{minipage}{0.01\linewidth} 
  \begin{center}
  \end{center}
 \end{minipage} 
 \begin{minipage}{.46\linewidth}
  \begin{center}
 \includegraphics[width=\linewidth]{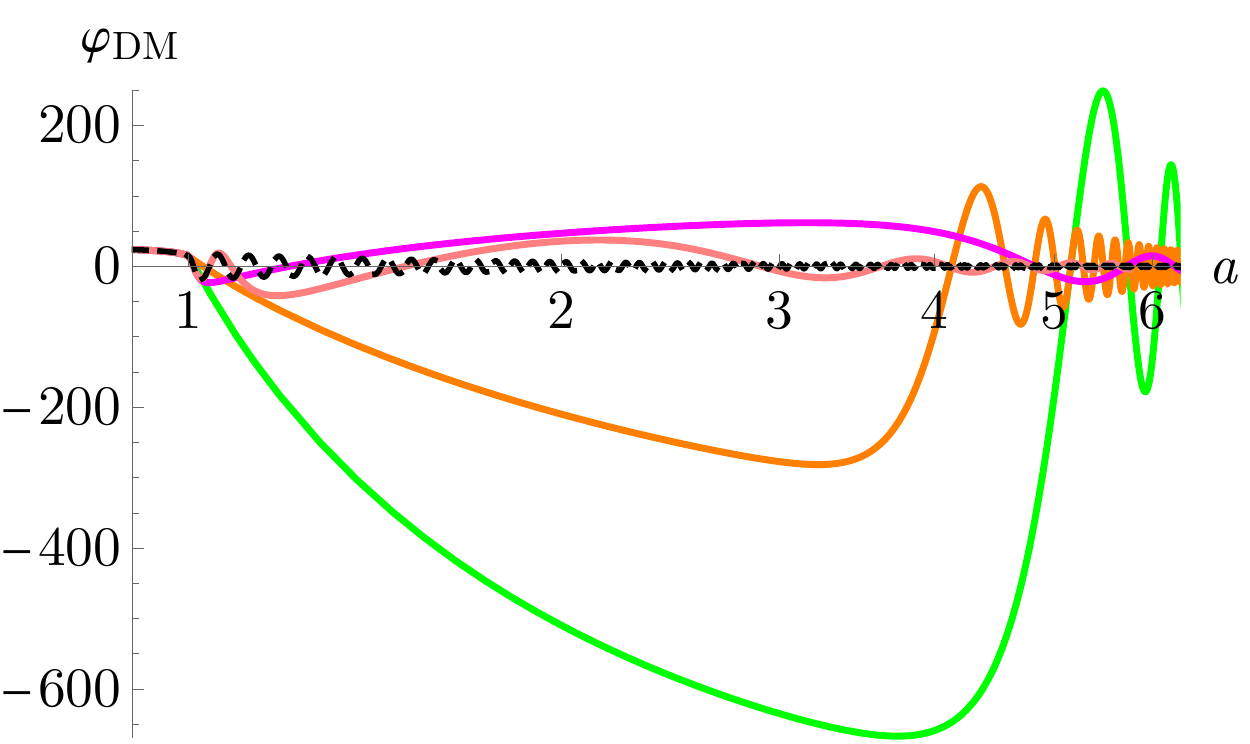}
  \end{center}
 \end{minipage} 

 \begin{minipage}{.46\linewidth}
  \begin{center}
 \includegraphics[width=\linewidth]{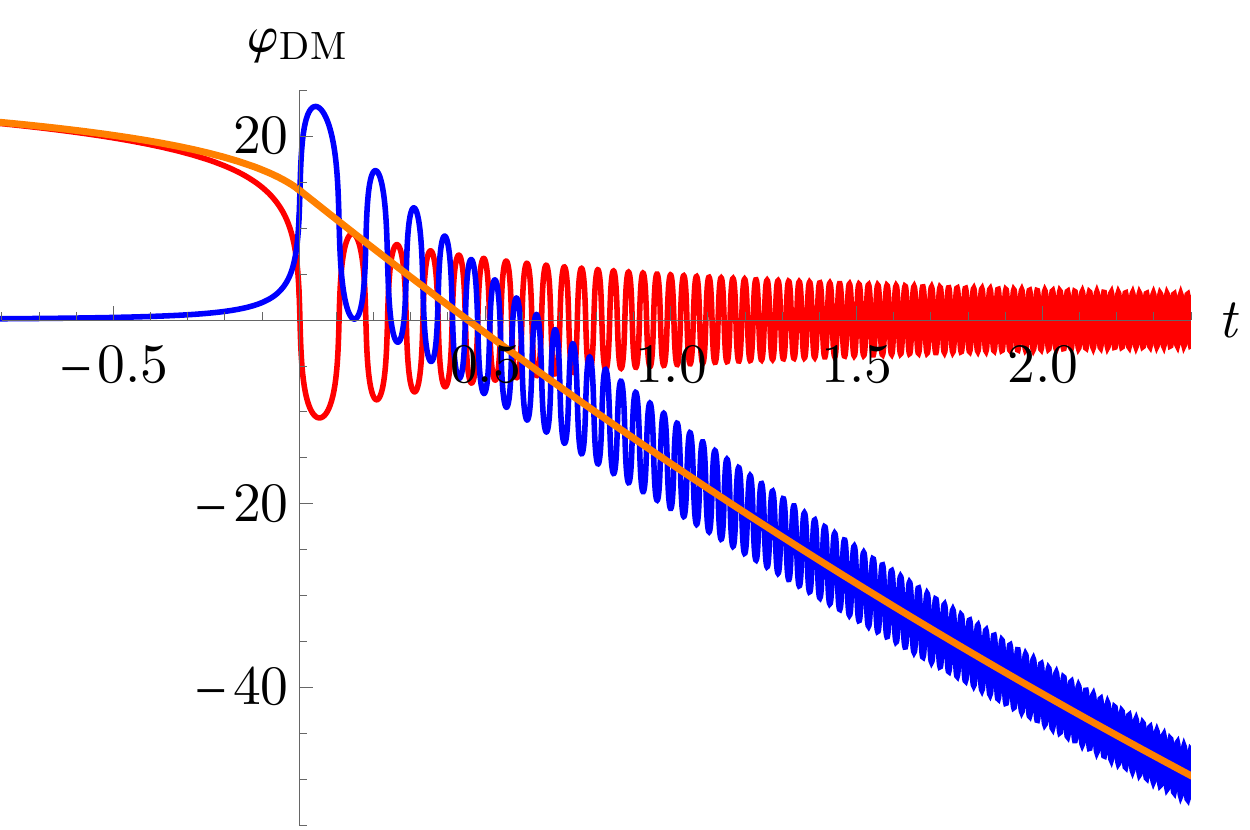}
  \end{center}
 \end{minipage} 
 \begin{minipage}{0.01\linewidth} 
  \begin{center}
  \end{center}
 \end{minipage} 
 \begin{minipage}{.46\linewidth}
  \begin{center}
 \includegraphics[width=\linewidth]{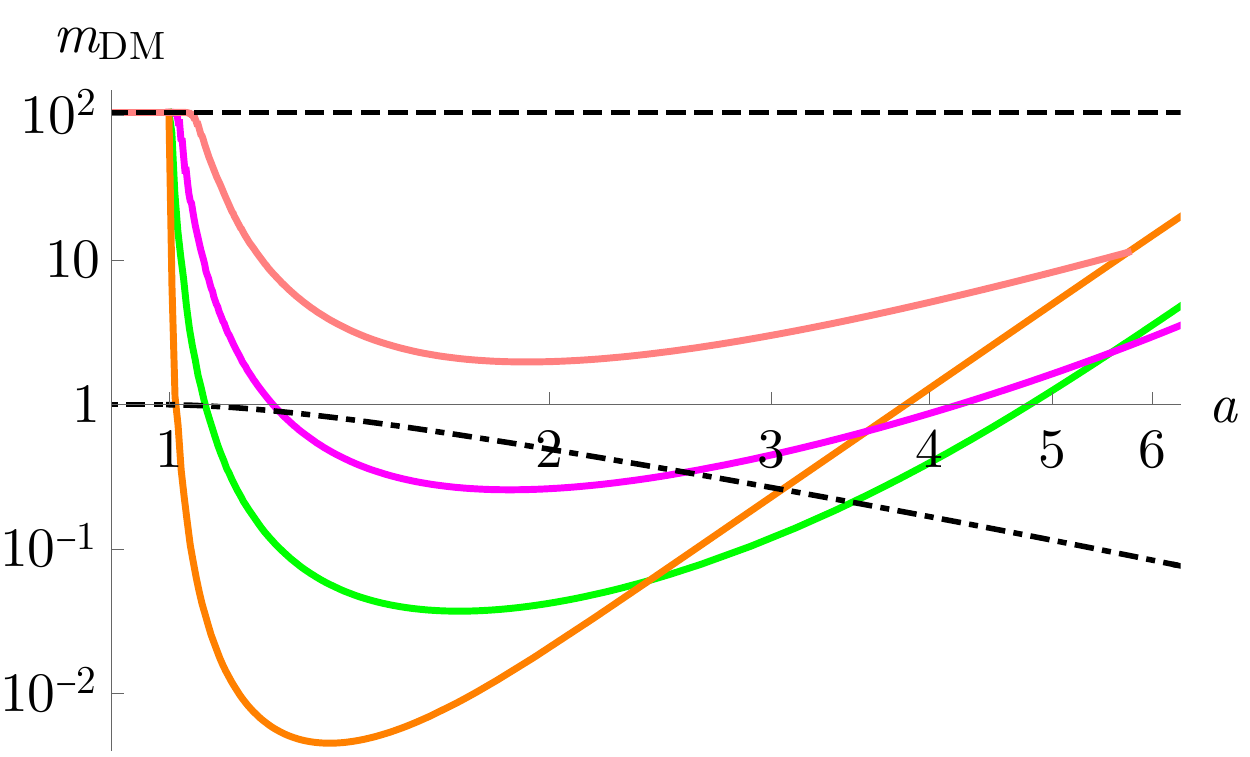}
  \end{center}
 \end{minipage} 
 \caption{Time evolution of the dark matter field and its mass for a
 case with an 
inflaton mass at the vacuum $m_{\phi 0} \approx 10^4 H_{\mathrm{end}}$,
zero-temperature axion mass $m_{\sigma 0} \approx 10^2
 H_{\mathrm{end}}$,
and strong coupling scale 
$\Lambda = 10^{-1} \cdot  (3 M_{\mathrm{Pl}}^2 H_{\mathrm{end}}^2)^{1/4} $.
The decay width of the reheaton is varied as 
$\Gamma_{\mathrm{RH}} / H_{\mathrm{end}} \approx 4$ (orange line), $0.7$
 (green), $0.2$ (magenta), $0.04$ (pink), and $0$ (black dashed).
The bottom left panel further shows $\sigma$ (blue) and $\alpha \phi$
 (red), and the bottom right panel shows~$H$ (black dot-dashed).
The value of $\varphi_{\mathrm{DM}}$ is
normalized by $\alpha M_{\mathrm{Pl}} H_{\mathrm{end}} /  m_{\phi 0}$, 
while $m_{\mathrm{DM}}$ is normalized by~$H_{\mathrm{end}}$.
Time~$t$ is in units of $2 \pi / m_{\sigma 0}$,
and the end of inflation is set to $ t_{\mathrm{end}}= 0$
and $ a_{\mathrm{end}}= 1$.
See the text for more details.}
 \label{fig:sim}
\end{center}
\end{figure}

The results of the numerical computations are displayed in 
Figure~\ref{fig:sim}, 
where the plots in the upper row show the evolution of the dark
matter field~$\varphi_{\mathrm{DM}}$ in terms of physical time
(upper left) and scale factor (upper right).
Here, the field value of~$\varphi_{\mathrm{DM}}$ is normalized by
$\alpha M_{\mathrm{Pl}} H_{\mathrm{end}} / m_{\phi 0}$,
and time~$t$ is in units of $2 \pi / m_{\sigma 0}$.
The end of inflation when $\dot{H} / H^2 = -1$ is set to
$t_{\mathrm{end}} = 0$ and $a_{\mathrm{end}} = 1$.
Each curve is plotted with a different value for the matter coupling:
$g_{\phi ff} = 0.1$ (orange), $0.04$ (green), $0.02$ (magenta), $0.01$
(pink), $0$ (black dashed).
The bottom left panel focuses on $g_{\phi ff} = 0.1$,
and shows the time evolution of 
$\sigma$ (blue) and $\alpha \phi$ (red), in addition to 
$\varphi_{\mathrm{DM}}$ (orange).
The normalization of the field values and time are the same as in the
upper row.
The bottom right panel shows the evolution of the dark matter
mass~$m_{\mathrm{DM}} $, 
which is approximately equal to the axion mass~$m_{\sigma}$, 
for each value of~$g_{\phi ff}$.
The Hubble rate~$H$ is also shown as the black dot-dashed line.
In this plot, the values of $m_{\mathrm{DM}}$ and $H$ are 
normalized by~$H_{\mathrm{end}}$.

The field evolution is, of course, independent of~$g_{\phi ff}$ until
the end of inflation, at which one sees that 
the dark matter field value is given by
(\ref{eq:varphi_DMend}) with $C \approx 14$. 
After inflation, a larger~$g_{\phi ff}$ gives a larger decay width
for the reheaton, and thus the radiation temperature rises more rapidly,
which in turn makes the axion mass decrease faster.
To understand the different behaviors for each value of~$g_{\phi ff}$,
it is instructive to see when the field evolution deviates from the 
case of $g_{\phi ff} = 0$ (black dashed) where there is no decay and
thus the dark matter field simply oscillates with frequency~$m_{\sigma
0}$.

For $g_{\phi ff} = 0.1$ (orange), the ratio between the
reheaton's decay width and the Hubble rate at the end of inflation is 
$\Gamma_{\mathrm{RH}} / H_{\mathrm{end}} \approx 4$.
The elapsed time~$\Delta t$ since the end of inflation until the axion mass
becomes smaller than the Hubble rate is
$\Delta t \, (m_{\sigma 0} / 2 \pi) \approx 0.2$,
namely, the time scale for the dark matter field to become effectively
massless is shorter than the initial oscillation period.
Consequently, the field begins to free-stream with 
an initial velocity $\simeq \dot{\varphi}_{\mathrm{DM\, end}} $.
The onset of the oscillation (i.e. when (\ref{c_pRD}) is satisfied)
is at $a_{\mathrm{osc}} \approx 3 a_{\mathrm{end}}$,
and the field displacement at this time 
is given by (\ref{eq:fin-dis}) with $B \approx 3$.
This example is well described by the simple picture outlined 
in Subsection~\ref{sec:kick}.

For $g_{\phi ff} = 0.04$ (green), the values become
$\Gamma_{\mathrm{RH}} / H_{\mathrm{end}} \approx 0.7$ and
$\Delta t \, (m_{\sigma 0} / 2 \pi) \approx 1$.
Here the dark matter field is slightly accelerated before starting to
free-stream, and thus the displacement 
at $a_{\mathrm{osc}} \approx 4 a_{\mathrm{end}}$
is enhanced to $B \approx 6 $.

For $g_{\phi ff} = 0.02$ (magenta), 
$\Gamma_{\mathrm{RH}} / H_{\mathrm{end}} \approx 0.2$ and
$\Delta t \, (m_{\sigma 0} / 2 \pi) \approx 4$.
The slowly-diminishing mass drags the dark matter field for a while and 
forces it to free-stream towards the positive direction in the plot.
However the amplitude of the initial velocity for the free-streaming is
still $\sim \abs{\dot{\varphi}_{\mathrm{DM\, end}}} $,
and thus the final displacement at 
$a_{\mathrm{osc}} \approx 4 a_{\mathrm{end}}$ becomes
$B \approx 0.5$.

For $g_{\phi ff} = 0.01$ (pink), $\Gamma_{\mathrm{RH}} /
H_{\mathrm{end}} \approx 0.04$, and
the mass never goes below the Hubble rate, hence the discussion
in the previous subsection does not directly apply. 
However, due to the mass becoming sufficiently smaller than its
zero-temperature value, the field dynamics is 
altered from the case with a constant mass.

In the bottom left plot,
the higher-frequency oscillation in the inflaton and axion fields
represents the reheaton degree of freedom.
After the reheaton decays away, the dark matter and axion fields become
approximately equivalent.

In the bottom right plot, a larger $g_{\phi ff}$ gives a 
smaller dark matter mass (axion mass) at the maximum
temperature~$T_{\mathrm{max}}$. 
After reaching its maximum value, the temperature drops as
$T \propto a^{-3/8}$ (cf. (\ref{eq:rad3over2})), and 
then as $T \propto a^{-1}$ after radiation domination takes over.
A larger $g_{\phi ff}$ yields an earlier radiation domination, 
which explains the faster growth of the mass seen in the plot.
The evolution of the Hubble rate also depends on~$g_{\phi ff}$, 
however since their differences in this log plot is insignificant, 
we have only shown the case for~$g_{\phi ff} = 0.02$. 

In the numerical examples presented here, all cases where the axion mass
becomes smaller than the Hubble rate (i.e. $g_{\phi ff} \geq 0.02$)
exhibit final dark matter field displacements with $B$ roughly of
order unity.
For these cases, 
the analytical arguments in Section~\ref{sec:kick} provide a good
effective description of the post-inflationary inflaxion dynamics.

\section{Parameter Space}
\label{sec:PM}

Let us put together the conditions for the temperature-dependent
inflaxion scenario. We have analyzed cases where the
cosmic temperature evolves as (\ref{ro4}),
with the axion and inflaton masses satisfying (\ref{hierarchy}).
The resulting axion dark matter abundance is given by
(\ref{omega-vm}), with the misalignment angle~(\ref{eq:theta_star}).
Upon deriving the abundance it was assumed that at the onset of the
axion oscillation, the universe is dominated by radiation,
i.e. (\ref{eq:dom-osc}),
and that the axion mass still has not reached its zero-temperature
value, i.e. (\ref{iiid}).
It was also assumed that the axion's self-interactions are
negligible, i.e. (\ref{eq:harmonic}),
and the present-day cosmic temperature is below the strong
coupling scale, $T_0 < \Lambda$. 
The cosmic temperature at its maximum is given in (\ref{T_max}),
the Hubble scale when radiation domination begins in (\ref{H_dom}),
and the temperature at the onset of the dark matter field oscillation 
in~(\ref{T_osc}).

A successful cosmology with this inflaxion scenario 
requires the reheaton to decay and trigger radiation domination at
temperatures of
$T_{\mathrm{dom}} \gtrsim 4\, \mathrm{MeV}$
so as not to spoil BBN~\cite{Kawasaki:2000en,Hannestad:2004px}, 
while the lifetime of the axion dark matter to be longer than the age of
the universe, 
$\Gamma_{\mathrm{DM}} < H_0 \approx 1 \times 10^{-33}\, \mathrm{eV}$.
($\Gamma_{\mathrm{DM}}$ depends on the cosmic temperature through 
$m_\sigma (T)$, and $\Gamma_{\mathrm{DM}} < H_0 $ should be satisfied
for the zero-temperature mass~$m_{\sigma 0}$. On the other hand
$\Gamma_{\mathrm{RH}}$ is almost independent of $m_\sigma (T)$
(cf. (\ref{eq:3.9})--(\ref{eq:3.11})), and thus is effectively constant
throughout the post-inflation epoch.)
Finally, the dark matter abundance should fulfill
$\Omega_\sigma h^2 \approx 0.1$ 
to match with observations.

\subsection{Case Study: $\sigma F \tilde{F} + \phi \bar{\psi} i \gamma^5 \psi$}

\begin{figure}[htbp]
\centering
\subfigure[$\Lambda = 200\, \mathrm{MeV}$ (QCD axion)]{%
  \includegraphics[width=.45\linewidth]{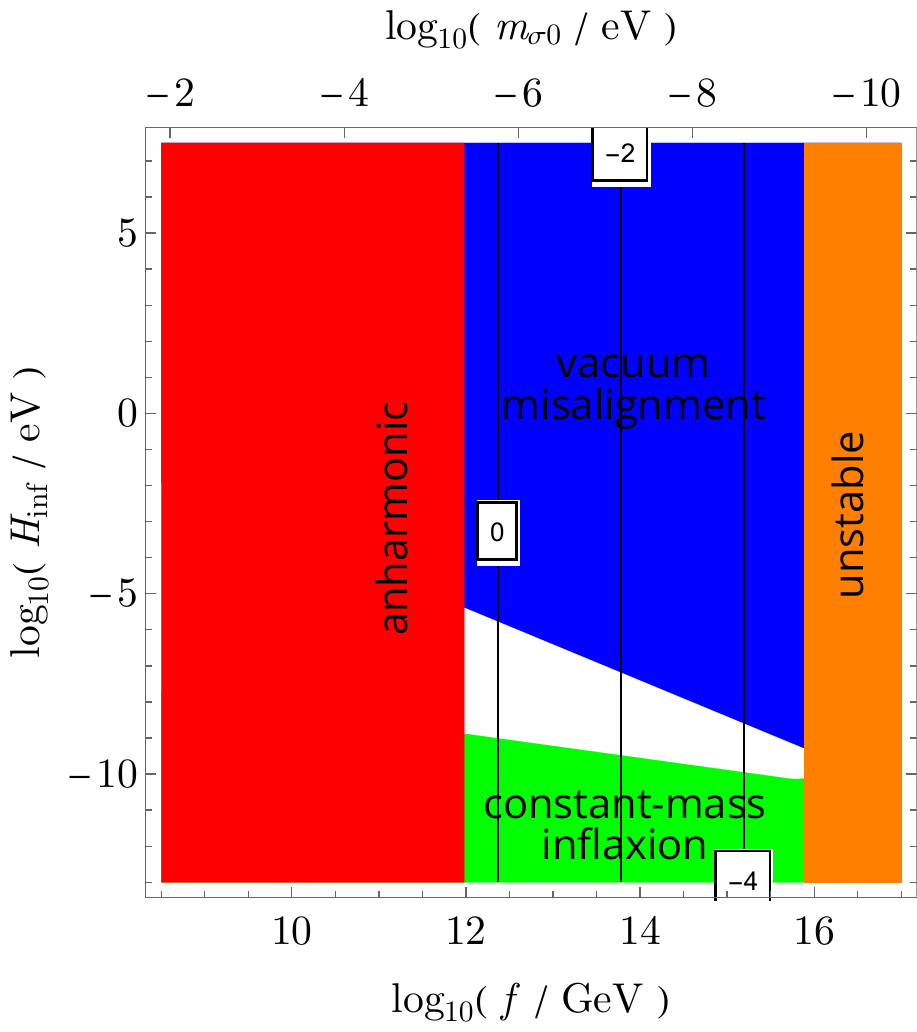}
  \label{fig:windowQCD}}
\quad
\subfigure[$\Lambda = 10\, \mathrm{GeV}$]{%
  \includegraphics[width=.45\linewidth]{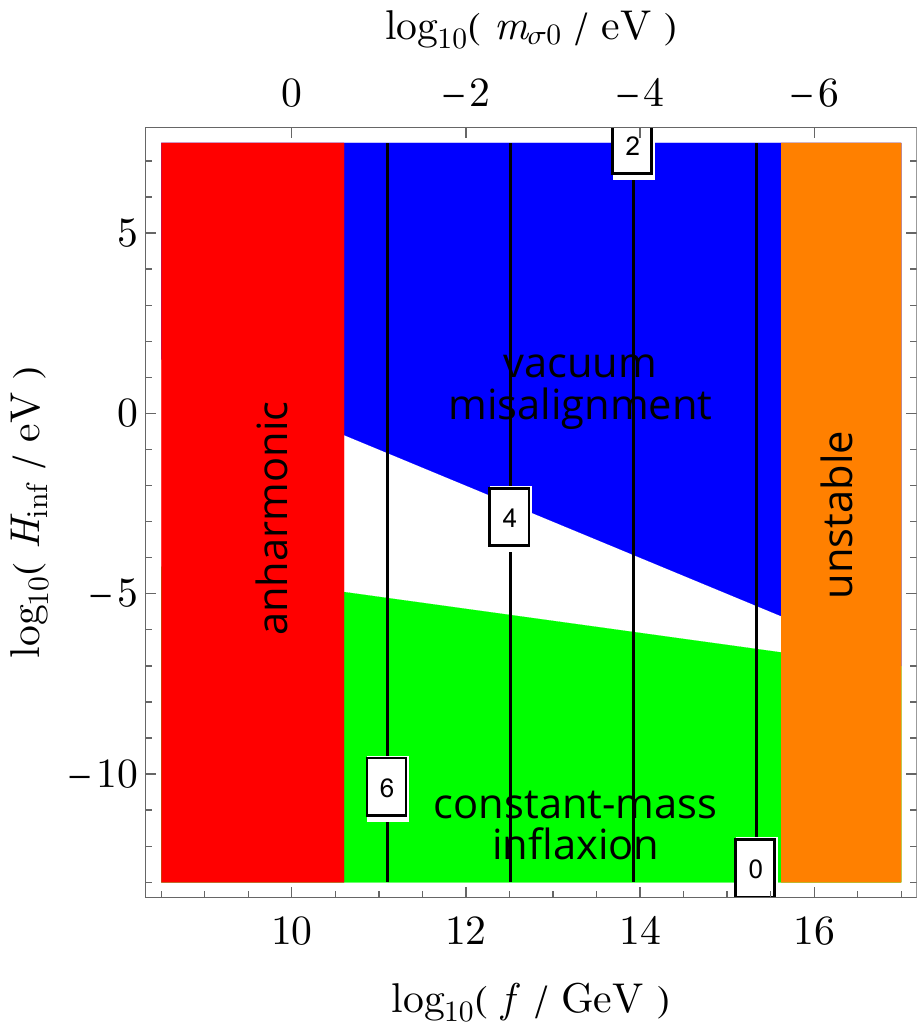}
  \label{fig:window1}}
\subfigure[$\Lambda = 10^3\, \mathrm{GeV}$]{%
  \includegraphics[width=.45\linewidth]{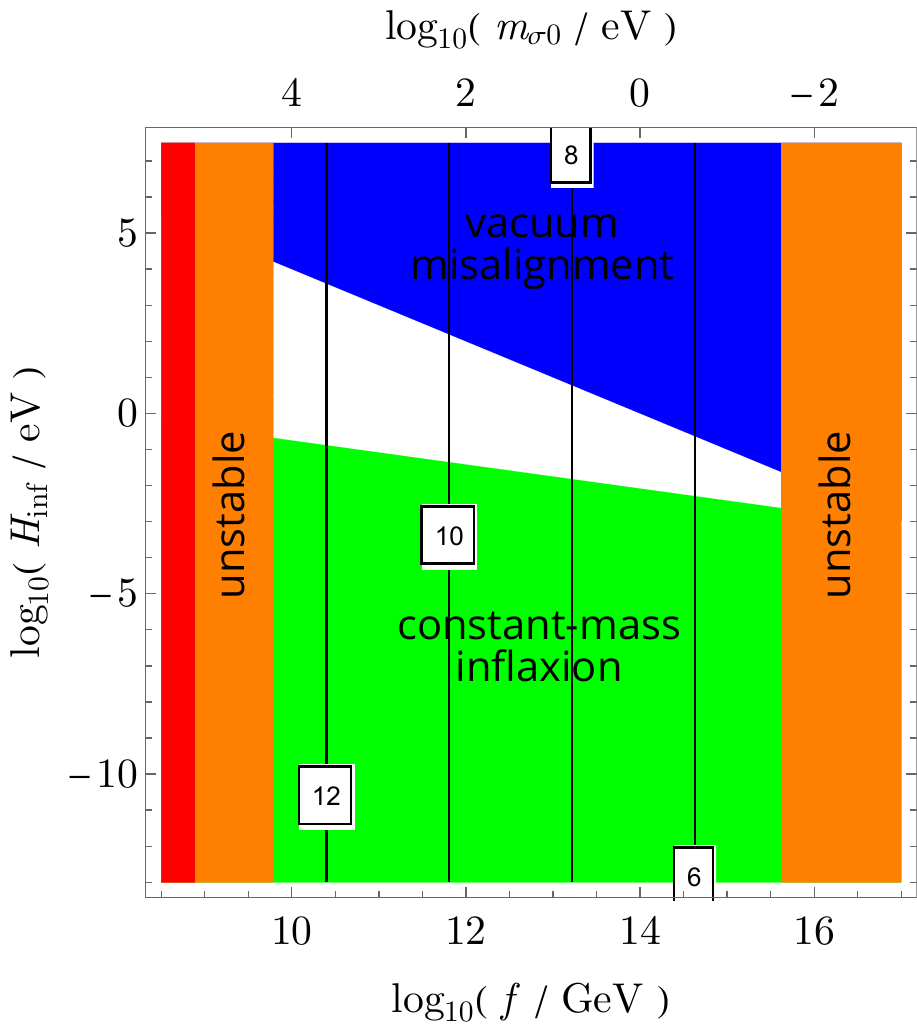}
  \label{fig:window3}}
\quad
\subfigure[$\Lambda = 10^5\, \mathrm{GeV}$]{%
  \includegraphics[width=.45\linewidth]{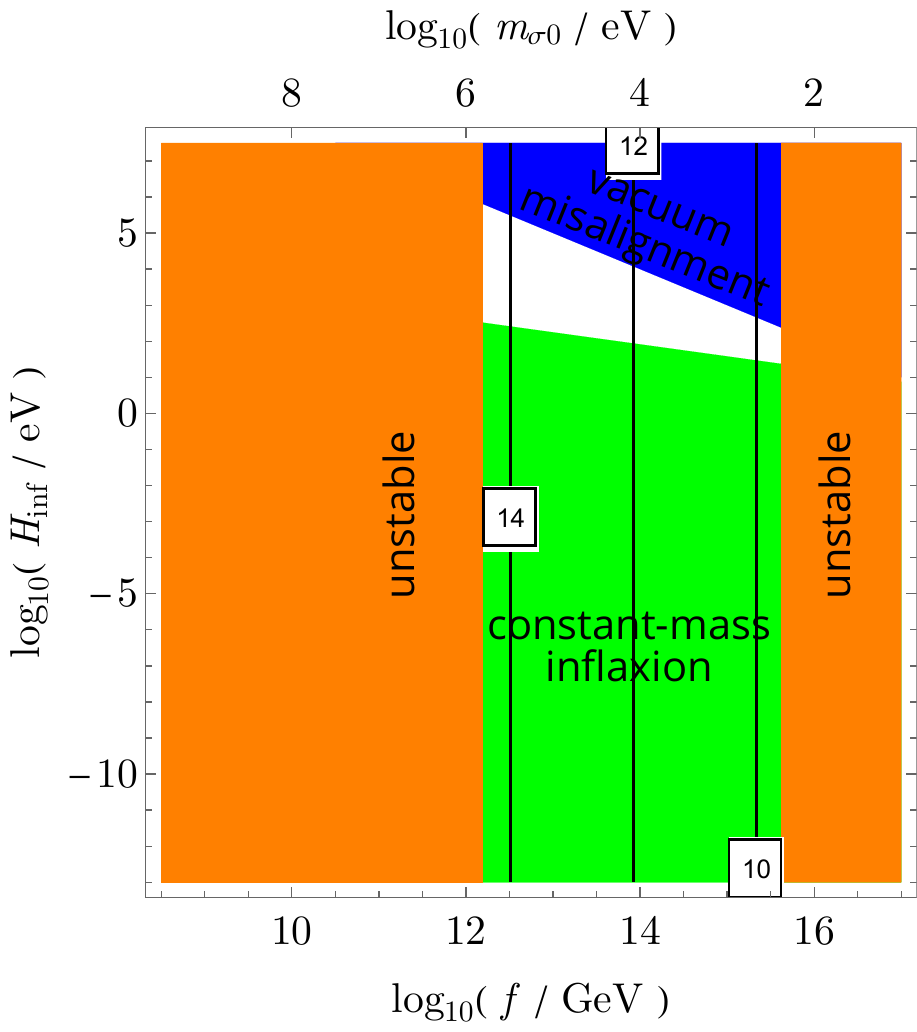}
  \label{fig:window5}}
 \caption{Parameter space for axion dark matter in the
 temperature-dependent inflaxion scenario, shown for the QCD axion and
 axion-like particles with different values of the strong coupling
 scale~$\Lambda$. 
 Axes are axion decay constant (bottom), 
 zero-temperature axion mass (top), 
 and inflationary Hubble scale (left). 
 The allowed windows are shown in white.
 Colored regions lie outside the validity of our analysis since the
 following conditions are violated: 
 $m_{\sigma 0} > H_{\mathrm{inf}}$ (blue),
 $m_\sigma (T_{\mathrm{max}}) < H_{\mathrm{inf}} $ (green),
 negligible axion self-interaction (red),
 and dark matter stability (orange).
 Contour lines show the inflaton mass at the vacuum in terms of
 $\log_{10} (m_{\phi 0} / \mathrm{eV})$. 
 The inflaton-axion mixing constant is set to $\alpha = 1/3$,
 and the inflaton is coupled to fermions with $g_{\phi ff} = 10^{-2}$. 
 See the text for more details.}  
\label{fig:inflaxion_window}
\end{figure}

In Figure~\ref{fig:inflaxion_window} we show the region of the axion
decay constant and inflation scale where all of the above conditions are
satisfied. 
(Here we denote the Hubble rate until the end of inflation collectively 
by~$H_{\mathrm{inf}}$, considering small-field inflation models in which
the time variation of the Hubble rate is tiny.)
We have fixed the axion-inflaton kinetic mixing constant to 
$\alpha = 1/3$,
and the axion mass parameters as 
$\xi  = 10^{-1}$, $\lambda = 10^{-1}$, and $p = 4$.
The parameter regions are shown for the QCD axion whose strong
coupling scale is $\Lambda = 200\, \mathrm{MeV}$, 
as well as for axion-like particles that obtain masses at higher energy
scales of $\Lambda = 10\, \mathrm{GeV}$, $10^3\, \mathrm{GeV}$, $10^5\,
\mathrm{GeV}$.
The value of the zero-temperature axion mass is shown in the upper axes.
The inflaton mass at the vacuum is fixed by the requirement of
$\Omega_\sigma h^2 \approx 0.1$,
and is shown on the contour lines in terms of
$\log_{10} (m_{\phi 0} / \mathrm{eV})$. 
For the matter couplings, we considered 
the axion to couple to SM/hidden photons, 
and the inflaton to light fermions via
\begin{equation}
 L_c = \frac{\alpha_{\gamma}}{8 \pi f} \sigma F_{\mu \nu} \tilde{F}^{\mu
  \nu}  + g_{\phi ff} \, \phi \bar{\psi} i \gamma^5 \psi.
\end{equation}
We took the dimensionless couplings as
$\alpha_{\gamma} = 10^{-2}$ and $g_{\phi ff} = 10^{-2}$,
and evaluated the decay widths as (\ref{eq:3.9}) and (\ref{eq:3.11}).
Moreover, the parameter~$B$ which characterizes
free-streaming (cf. (\ref{eq:fin-dis})) was taken as $B = 1$.

The regions where the temperature-dependent inflaxion scenario allows
for a successful reheating and axion dark matter generation are shown in
white. On the other hand in the colored regions, the conditions listed
above are violated.
For the chosen set of parameters, there are four conditions that most
severely constrain the parameter space:
The blue region violates $m_{\sigma 0} > H_{\mathrm{inf}}$; this is the
region where the conventional vacuum misalignment scenario can operate. 
The green region violates 
$m_\sigma (T_{\mathrm{max}}) < H_{\mathrm{inf}} $
and thus lies outside the validity of the analysis in this paper.
Deep inside this region the axion mass would stick to its
zero-temperature value throughout the reheating process; 
here dark matter can instead be produced by the constant-mass 
inflaxion scenario as discussed in~\cite{Kobayashi:2019eyg}. 
The red region violates $\abs{\theta_\star} < 1$;
here the axion cannot account for all of dark matter
without invoking anharmonic effects, as we already saw in
Figure~\ref{fig:kappa_p}. 
Within this region, the inflaxion scenario can produce the observed dark
matter abundance only if the axion field after free-streaming 
happens to land in the vicinity of a potential maximum.\footnote{Cases
with $\abs{\varphi_{\mathrm{DM} \star}} > \pi f$ which typically give
$\abs{\theta_\star} \sim 1$
(as discussed below (\ref{eq:harmonic}))
live on the edge of the red region.}
The orange region violates $\Gamma_{\mathrm{DM}}  < H_0$  and
thus dark matter would not survive until today.
In the orange regions at 
small~$f$ (close to the left edges of the plots)
the dark matter decays predominantly through the axion-photon coupling,
while in the regions at large~$f$ (close to the right edges) 
the decay is via the inflaton-fermion coupling.
The orange exclusion regions at large~$f$ disappear if the fermion
mass lies within 
$m_{\sigma 0} < 2 m_{\psi} < m_{\mathrm{RH}}$ and thus the decay of dark
matter into fermions is kinetically forbidden; in such a case the
allowed windows extend to even larger $f$ values until they hit
other conditions such as (\ref{iiid}).
The reheaton decays through the
inflaton-fermion coupling in all four plots.

We also note that for the chosen set of parameters, 
radiation domination takes over shortly after the end of inflation
(i.e. $\Gamma_\mathrm{RH} > H_{\mathrm{end}}$)
in all of the allowed windows, except for in the vicinity of the
upper right corner of the windows of Figures~\ref{fig:windowQCD} and
\ref{fig:window1}.

The inflaxion mechanism can also operate with
other forms of the matter couplings, for instance, with an
inflaton-photon interaction $\phi F\tilde{F}$.  
However with only the $\sigma F \tilde{F}$ coupling and no matter
couplings for the inflaton, 
there is no parameter space that satisfies all
the conditions, independently of the value of the 
coupling strength~$G_{\sigma \gamma \gamma}$,
if the other parameters take the same values as in the above example.

\subsection{Upper Bound on Inflaton Mass}

Independently of the details of the matter couplings, the inflaton mass
in this scenario is bounded from above as follows.
For simplicity, let us here take 
$\kappa_p \sim 0.1$,
$g_{(s)*} (T_{\mathrm{osc}}) \sim 100$, 
$\lambda \sim 0.1$,
$\xi \sim 0.1$
in the expression for the relic abundance (\ref{omega-vm}).
Then using (\ref{eq:theta_star}) with $B \sim 1$ for the misalignment
angle, the normalization of $\Omega_\sigma h^2 \approx 0.1$ fixes
the inflaton mass as 
\begin{equation}
 m_{\phi 0} \sim 10\, \mathrm{eV} \cdot
\abs{\alpha} \left( \frac{\Lambda}{200 \,
	      \mathrm{MeV}}\right)^{\frac{5}{2}} 
\left(\frac{f}{10^{12}\, \mathrm{GeV}} \right)^{-\frac{3p+5}{2 p+4}}.
\label{eq:4.2}
\end{equation}
On the other hand, the requirement of $\abs{\theta_\star} < 1$ 
under $\Omega_\sigma h^2 \approx 0.1$ 
bounds the decay constant as
\begin{equation}
 f \gtrsim 10^{12}\, \mathrm{GeV}
\left( \frac{\Lambda }{200 \, \mathrm{MeV}} \right)^{-\frac{p+2}{p+3}},
\end{equation}
as shown by the red regions in the plots. 
Combining these expressions yields a bound on the inflaton
mass that is independent of the matter coupling,
\begin{equation}
 m_{\phi 0} \lesssim
10 \, \mathrm{eV} \cdot \abs{\alpha}
\left( \frac{\Lambda}{200\, \mathrm{MeV}} \right)^{\frac{4 p+10}{p+3}}.
\label{eq:mphi-bound}
\end{equation}
This actually sets the upper bounds on the inflaton mass in 
Figures~\ref{fig:windowQCD} and \ref{fig:window1},
while in Figures~\ref{fig:window3} and \ref{fig:window5} the dark matter
stability condition gives stronger bounds.
For axion-like particles coupled to a 
new strong gauge group (not QCD) with a confinement scale $\Lambda
\gg 200\, \mathrm{MeV}$, the upper limit 
(\ref{eq:mphi-bound}) allows for a heavy enough inflaton
so that perturbative reheating is easy to implement successfully.


\subsection{QCD Inflaxion}

For the QCD axion, the bound (\ref{eq:mphi-bound}) is particularly
restrictive, which together with $\abs{\alpha} < 1$ gives
$m_{\phi 0 } \lesssim 10 \, \mathrm{eV}$.
This rather small inflaton mass, and hence a small reheaton mass (unless
$\abs{\alpha}$ is very close to unity), poses a challenge for 
perturbative reheating.\footnote{The QCD inflaxion scenario in which the
temperature of the Universe never reaches values above $\Lambda$ also
has a similar issue; see the appendix in Ref.~\cite{Kobayashi:2019eyg}.}
The only SM states kinematically
accessible are photons and neutrinos
(although not necessarily all three neutrinos, depending on the value of 
$m_{\phi 0 }$).
Reheating above the BBN temperature by decaying into photons 
requires an operator of the sort $\phi F \tilde{F} $ with an extremely
large coupling strength for $m_{\phi 0} \lesssim 10$ eV,
such that it is largely excluded by stellar cooling
bounds~\cite{Tanabashi:2018oca}. 
It would be interesting to study the non-perturbative preheating
phase with the photon coupling
(see e.g. \cite{Adshead:2015pva,Cuissa:2018oiw})
 to assess whether it is a viable option,
but that is beyond the scope of the current work.

We consider then the decay into neutrinos. 
Gauge invariance of the SM dictates that the lowest-dimensional
operator available is of dimension six:
\begin{equation} \label{op6}
\frac{Y^{ij}}{\Lambda_6^2} \phi (\epsilon^{ab} {\mathbf H}_a {\mathbf L}_{bi})(\epsilon^{cd} {\bf H}_c {\bf L}_{dj}) + {\rm h.c.}
\end{equation}
Here ${\bf H}$ and ${\bf L}$ are the Higgs and lepton SM fields, respectively; $a,b,c,d$ are $SU(2)_L$ indices, while 
$i,j$ flavor indices; $Y^{ij}$ are generalized yukawa couplings. 
We write this operator using two-component spinor notation, following the conventions
of Ref.~\cite{Dreiner:2008tw}. 
In the scenario under consideration the electroweak symmetry is 
broken throughout the cosmic history (cf. Figure~\ref{fig:windowQCD}), 
and thus the operator gives rise to a yukawa coupling of the inflaton
to the left-handed SM neutrinos,
\begin{equation} \label{yuknu}
y_6^{ij} \phi \nu_i \nu_j \, , \qquad y_6^{ij} = Y^{ij} \frac{v^2}{\Lambda_6^2} \, ,
\end{equation}
where $v$ is the electroweak scale.
From here on we drop the flavor indices, for the sake of brevity, 
and we take the entries of $Y^{ij}$ to be of order one. 
The interaction \eqref{yuknu} leads to a decay rate of the reheaton into neutrinos
that has the form of \eqref{eq:3.11} with $g_{\phi ff}$ replaced by $y_6$. 
The working assumption~(\ref{eq:dom-osc}) adopted in this
paper\footnote{We have also considered the possibility
of $T_{\rm dom} \ll \Lambda < T_{\rm max}$, 
so that the dark matter field
begins to oscillate before entering radiation domination.
Here the relic abundance becomes different from
the one we reviewed in Section~\ref{sec:vm},
and it depends also on $T_{\rm dom}$~\cite{Giudice:2000ex} . 
However we found that the parameter window for this case 
is tiny in our inflaxion scenario.
The reason is that the following three conditions:
(i) getting the observed dark matter abundance, 
(ii) having $H_{\rm inf} < m_{\sigma 0}$,
and (iii) $T_{\rm dom} \ll T_{\rm max}$,
are incompatible with each other in most of the parameter space.}
requires $T_{\mathrm{dom}}$ to be above $\Lambda \approx 200$ MeV,
which implies 
\begin{equation} \label{yLambda}
y_6 > 10^{-5} \, , \qquad \Lambda_6 < 10^{5/2} v \simeq 80 \ {\rm TeV} \, ,
\end{equation}
with $m_{\phi 0} = 10$ eV. A few comments are in order.

At the beginning of the oscillating phase the reheaton field describes a collection
of non-relativistic scalar particles, which decay to produce neutrinos. These,
in turn, interact among themselves via the weak force to quickly populate and 
thermalize the SM sector. The coupling \eqref{yuknu} also implies that the scattering
rate involving neutrinos and the reheaton remains faster than the Hubble expansion rate
as the temperature decreases. Therefore the $\varphi_{\mathrm{RH}}$ quanta are upscattered and 
remain in the thermal bath with the neutrinos. 
This scenario, to our knowledge, has not been explored in detail yet
and we leave a dedicated study of its cosmological implications to
future work.\footnote{Such a case is also touched upon in the conclusions
of Ref.~\cite{Grohs:2020xxd}, and it could also have implications in
addressing the $H_0$ tension \cite{Blinov:2019gcj}.}

On the particle physics side there are many constraints to take into account. 
First, we note that the operator \eqref{op6} does not directly
contribute to neutrino masses, given that the vacuum expectation value
(VEV) of the inflaton approaches zero.
Second, by replacing one ${\bf H}$ with its VEV, it leads to the three body decay
${\bf H} \to \phi \nu\nu$; the corresponding width, for $\Lambda_6$ 
not too far from the upper bound \eqref{yLambda}, 
is very small and the bound from invisible Higgs decays is amply evaded.
Third, we note that the operator
\eqref{op6} contains also charged leptons which, due to the $SU(2)_L$ structure, are always
accompanied by a charged Higgs, that becomes the longitudinal mode of the $W$ boson after
electroweak symmetry breaking. The presence of the heavy $W$ bosons, in combination with
the suppression scale of \eqref{yLambda}, makes it hard to probe
our operator at colliders like LEP or LHC.
Fourth, lepton flavor violating processes are likely to constrain some of the entries of $Y^{ij}$ in \eqref{op6},
but unlikely to exclude completely our scenario. 
We reserve a more detailed study of the experimental constraints for the future.

\section{Conclusions}
\label{sec:conc}

A kinetic mixing between the axion and the inflaton can induce axion
dark matter production even if the inflationary Hubble scale is smaller
than the zero-temperature axion mass. 
Together with our previous analysis~\cite{Kobayashi:2019eyg}, 
we have explored two production scenarios for axions coupled to a strong
gauge group within this inflaxion framework where
(1) the reheating scale is lower than the strong coupling
scale and thus the axion mass stays constant throughout the
cosmic history, and 
(2) the maximum temperature during reheating exceeds the strong coupling
scale such that the axion mass temporarily vanishes.
The main part of this paper was devoted to case~(2), for which we 
found that the axion gets kicked out of the vacuum towards
the end of inflation, and subsequently in the reheating epoch 
drifts away even further from its potential minimum. 
The field dynamics during reheating thus 
gives rise to a misalignment angle which
sources axion dark matter in the later universe.
This `initial' misalignment angle is uniquely fixed by the Lagrangian
parameters as (\ref{eq:theta_star}), which is in contrast to the initial
angle in the conventional vacuum misalignment scenario being a random
variable. 
Our scenario further opens up new parameter space for axion dark matter,
in particular the regions with low inflation scales and large axion
decay constants.

Producing the QCD axion within our model requires careful consideration
of the reheating process due to the small inflaton mass, $m_{\phi 0}
\lesssim 10\, \mathrm{eV}$,
which is needed to obtain the observed dark matter abundance. 
We found that reheating into Standard Model neutrinos is an option,
which may also yield experimentally accessible new phenomena,
although we reserve a detailed study of the constraints 
on this scenario for the future. 
For axion-like particles coupled to a 
hidden strong gauge group that confines at an energy
$\Lambda \gg 1 \, \mathrm{GeV}$, all the scales involved (including the
inflaton mass) are
higher compared to the QCD case, 
and thus there are many possibilities for the reheating process.

Perhaps the most exciting feature of the inflaxion framework is the
inevitable link between the reheating temperature and the coupling of the
axion to normal matter, which is induced by the inflaton-axion kinetic
mixing. 
This could offer the possibility of probing the reheating scale
with laboratory experiments for measuring axion couplings, 
and/or astrophysical experiments for constraining the dark matter
lifetime. 
We also remark that, while this work mainly focused on the
homogeneous evolution of the inflaton-axion system,
depending on the form of the inflaton potential, inhomogeneities can
develop around the end of inflation. This may give rise to axion dark
matter clumps, which would further provide 
observational opportunities.

\section*{Acknowledgments}
We thank Francesco D'Eramo, Herbi Dreiner, Juan Herrero-Garcia, Kazunori
Kohri, Diego Redigolo, Andrea Romanino, and Yuko Urakawa for useful
discussions. 


\appendix

\section{Onset of Axion Oscillation}
\label{app:onset}

In this appendix we analyze when an axion with a temperature-dependent
mass starts to oscillate in the early universe as the mass becomes
larger than the Hubble rate.
Following the discussion in Appendix~A of~\cite{Kobayashi:2017jcf},
we explicitly solve the homogeneous Klein-Gordon equation 
in a flat FRW background,
\begin{equation}
 \ddot{\sigma} + 3 H \dot{\sigma} + m_\sigma^2 \sigma = 0.
\label{KG}
\end{equation}
We consider the background universe to have a constant equation of
state parameter~$w$ lying within the range $-1 < w < 1$, 
so that 
\begin{equation}
 H \propto a^{-\frac{3 (1+w)}{2}}.
\end{equation}
We also assume the axion mass to have a power-law dependence on the 
scale factor, 
\begin{equation}
 m_\sigma \propto a^p,
\end{equation}
with $p \geq 0$.
This corresponds to a temperature dependence of $m_\sigma \propto
T^{-p}$, given that the temperature scales as $ T \propto a^{-1}$. 
Furthermore, we assume the axion in the asymptotic past 
to be frozen at some field value~$\sigma_\star$.
Then the solution of (\ref{KG}) satisfying such an initial condition 
is obtained as
\begin{equation}
 \sigma = \sigma_\star \, \Gamma (\nu + 1)
\left( \frac{z}{2}  \right)^{-\nu} 
J_\nu (z),
\end{equation}
where $J_\nu (z)$ is the Bessel function of the first kind, and 
\begin{equation}
 z = \frac{ 2}{3 (1+w) + 2p}  \frac{m_\sigma }{H}, 
\quad
 \nu = \frac{3 (1-w)}{6 (1+w) + 4 p}.
\end{equation}
Using this exact solution,
the physical number density of the axion is computed as
\begin{equation}
n_\sigma =  \frac{1}{2 m_\sigma}
\left( \dot{\sigma}^2 + m_\sigma^2 \sigma^2 \right) 
 = \frac{m_\sigma \sigma_\star^2 }{2}  \, 
\Gamma (\nu + 1)^2
\left(\frac{z}{2}\right)^{-2 \nu}
\left\{
J_{\nu + 1} (z)^2 + J_{\nu } (z)^2
\right\}.
\label{n-exact}
\end{equation}

In the asymptotic past $ z \to 0$, the expression~(\ref{n-exact})
takes a limiting form of
\begin{equation}
 n_\sigma \sim \frac{1}{2} m_\sigma \sigma_\star^2,
\label{as_past}
\end{equation}
as set by the initial condition.
On the other hand, in the asymptotic future $z \to \infty$, it approaches
\begin{equation}
 n_\sigma \sim
\frac{m_\sigma \sigma_\star^2}{2 \pi } \, 
\Gamma (\nu+1)^2 \left( \frac{z}{2}  \right)^{-1-2 \nu }
\propto a^{-3},
\label{as_future}
\end{equation}
which manifests the conservation of the comoving number density
$n_\sigma a^3$.

\begin{figure}[t]
\begin{center}
 \begin{minipage}{.46\linewidth}
  \begin{center}
 \includegraphics[width=\linewidth]{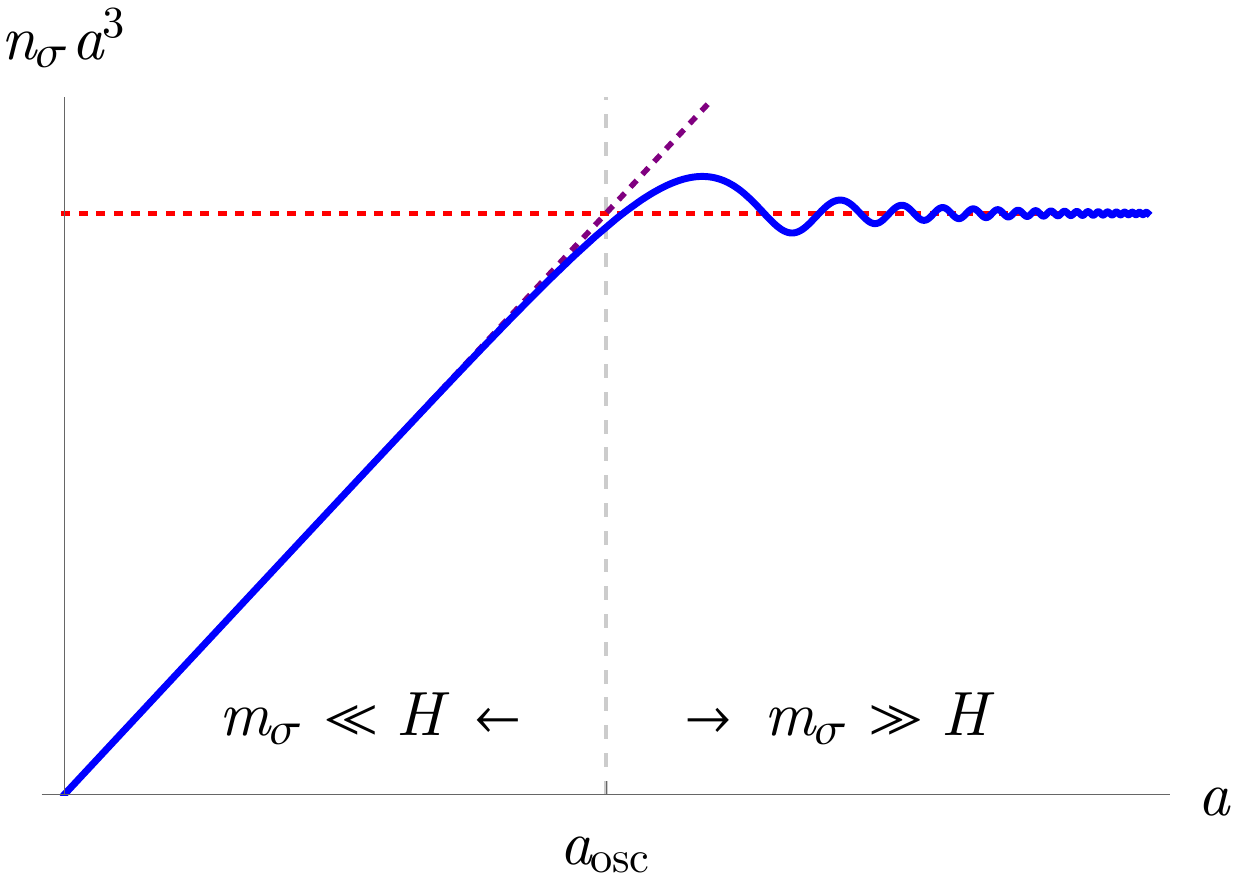}
  \end{center}
 \end{minipage} 
 \begin{minipage}{0.01\linewidth} 
  \begin{center}
  \end{center}
 \end{minipage} 
 \begin{minipage}{.46\linewidth}
  \begin{center}
 \includegraphics[width=\linewidth]{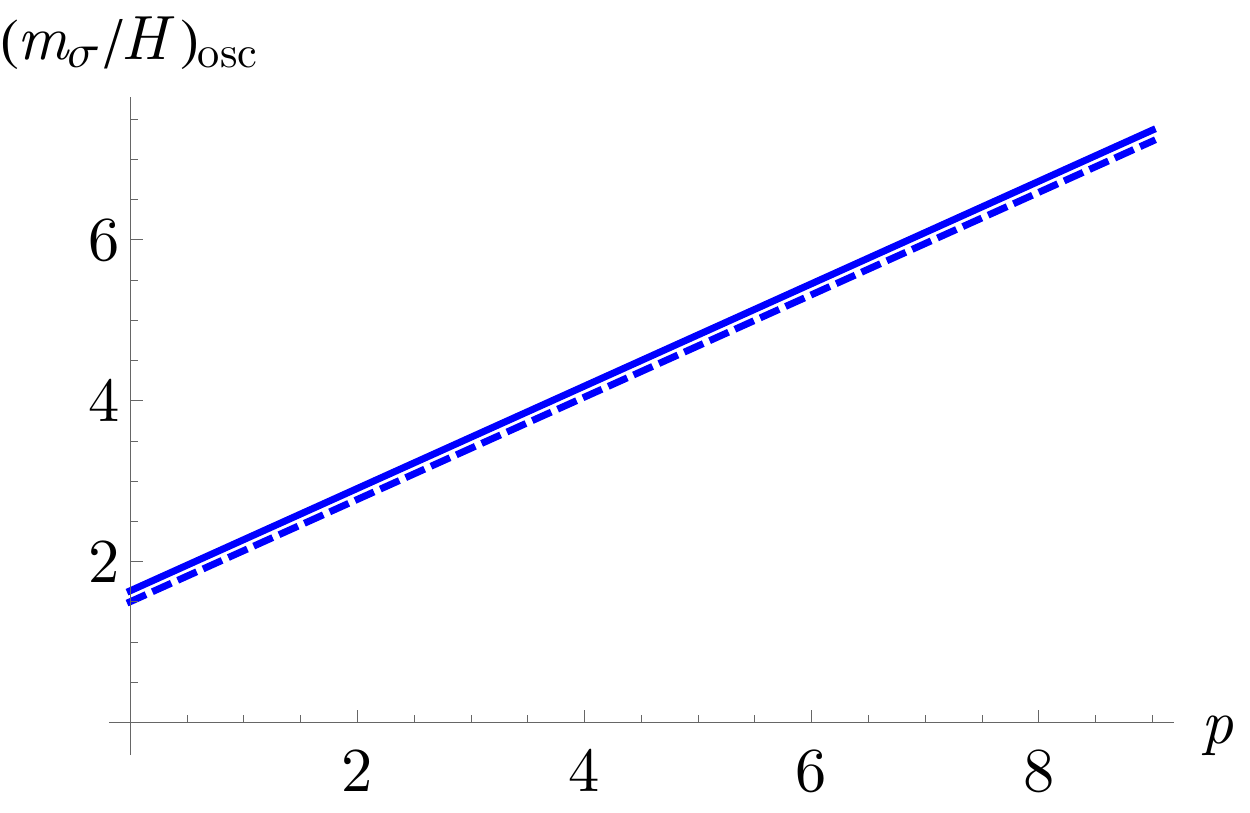}
  \end{center}
 \end{minipage} 
 \caption{Left: Time evolution of the axion's comoving number density in
 terms of the scale factor, in log-log scale. 
 The dotted lines show the limiting behaviors in the asymptotic past and
 future. The onset of the axion oscillation is defined as when the
 limiting forms cross each other (see the text for details). 
 Right: Ratio between the axion mass and the Hubble scale at
 the onset of the axion oscillation, as a function of the
 power~$p$ of the temperature dependence. 
 The equation of state of the universe is taken as $w = 1/3$ (solid
 line) and $w = 0$ (dashed).}
 \label{fig:mH_osc}
\end{center}
\end{figure}

The comoving number density as a function of the scale factor is 
illustrated in the left panel of Figure~\ref{fig:mH_osc} in a log-log plot.
The exact solution~(\ref{n-exact}) is represented by the blue solid
line, while the limiting forms in the asymptotic past~(\ref{as_past})
and future~(\ref{as_future}) are shown respectively as the purple and
red dotted lines.
Extrapolating these two limiting expressions to all times, 
one finds that they cross each other when the 
ratio between the axion mass and Hubble scale becomes
\begin{equation}
 \left( \frac{m_\sigma}{H} \right)_{\mathrm{osc}}
=  \frac{3 (1-w)}{2 \nu }
\left[
\frac{\Gamma (\nu + 1)^2}{\pi}
\right]^{\frac{1}{2 \nu + 1}}.
\label{eq:mH_osc}
\end{equation}
We refer to this time as the `onset' of the axion oscillation, and
denote quantities at this time by the subscript~$(\mathrm{osc})$.
This definition allows one to rewrite the conserved
comoving number density in the asymptotic future 
in terms of quantities at the onset of the oscillation as
\begin{equation}
 \lim_{t \to \infty} (n_\sigma a^3) = 
\frac{1}{2} m_{\sigma \mathrm{osc}} \, \sigma_\star^2
a_{\mathrm{osc}}^3,
\end{equation}
which is convenient for computing the relic abundance
as discussed in the main part of the paper.

The mass-to-Hubble ratio~(\ref{eq:mH_osc}) at the onset of the
oscillation is plotted against~$p$ in
the right panel of Figure~\ref{fig:mH_osc}
for $w = 1/3$ and $0$.
For instance, the QCD axion possesses 
a temperature-dependent mass of 
$m_\sigma \propto T^{-p} \propto a^{p}$ with $p\approx 4$
at $T \gg 200 \, \mathrm{MeV}$;
given that it starts to oscillate at such temperatures
during the radiation-dominated epoch,\footnote{The time variation of the
relativistic degrees of freedom~$g_*$ gives corrections to the scaling relation
$T \propto a^{-1}$ as well as to the equation of state $w =
1/3$. However as long as $g_*$ stays constant while the axion makes the
transition from the frozen to the oscillatory phase, the discussion
here is valid.}
i.e. $w = 1/3$, the mass-to-Hubble ratio is 
$(m_\sigma /H)_{\mathrm{osc}} \approx 4$.
We also remark that for an axion-like particle whose mass depends
sensitively on the temperature such that $p \gg 1$,
the ratio becomes as large as $(m_\sigma
/H)_{\mathrm{osc}} \gg 1 $. 
In such cases, computing the relic abundance based on 
a naive guess of $(m_\sigma /H)_{\mathrm{osc}} \sim 1 $ would lead to
quite inaccurate results.

\section{Diagonal Basis}
\label{app:diagonal}

We list expressions regarding the diagonal basis of the
inflaxion Lagrangian~(\ref{eq:Lagrangian}).
In this appendix we split the inflaton potential into a quadratic part
and the rest as 
\begin{equation}
 V(\phi) = \frac{1}{2} m_{\phi 0}^2 \phi^2 + U_{\mathrm{int}}(\phi).
\end{equation}
If the temperature dependence of the axion mass~$m_\sigma$ can be
ignored, the quadratic terms can be diagonalized and the Lagrangian can
be rewritten as 
\begin{equation}
 \frac{\mathcal{L}}{\sqrt{-g} } = 
\sum_{i=\pm} \left(
-\frac{1}{2} g^{\mu \nu} \partial_\mu \varphi_i \, \partial_\nu \varphi_i
- \frac{1}{2} m_i^2 \varphi_i^2 
\right)
- U_{\mathrm{int}}(\phi)
+ L_{\mathrm{c}}[\sigma, \phi, \Psi],
\label{diagonal_L}
\end{equation}
in terms of the diagonal fields and their masses:
\begin{equation}
 \varphi_{\pm} = 
\left(\frac{X_{\pm} +  \beta^2}{ X_{\pm}^2 + \alpha^2 \beta^2 }\right)^{1/2}
\left(  X_{\pm} \, \sigma + \alpha \, \phi \right),
\quad
 m_{\pm}^2 = \frac{1 -  X_{\pm} }{1 - \alpha^2} \, m_{\phi 0}^2,
\label{varphi_pm}
\end{equation}
where
\begin{equation}
 X_{\pm} =  \frac{ 1 -\beta^2 \pm \sqrt{ (1 - \beta^2)^2 + 4 \alpha^2
 \beta^2 }}{2},
\quad
 \beta^2 = \frac{m_\sigma^2}{m_{\phi 0}^2}.
\end{equation}

The diagonal field basis is also convenient for computing the 
decay widths of the scalar particles. 
If for instance we take the matter couplings as~(\ref{eq:couplings}),
then by rewriting $(\phi, \sigma)$ in terms 
of $(\varphi_+, \varphi_-)$, the decay widths of the diagonal
fields are obtained as
\begin{align}
  \Gamma (\varphi_\pm \to \gamma \gamma)
& 
\simeq 
\frac{G_{\pm \gamma \gamma }^2 }{64 \pi } m_\pm^3,
&
 G_{\pm \gamma \gamma} 
& 
= 
\frac{1}{X_+ - X_-}
 \left(\frac{X_{\pm} +  \beta^2}{ X_{\pm}^2 + \alpha^2 \beta^2
  }\right)^{-1/2}
\left(
G_{\sigma \gamma \gamma} - \frac{X_\mp}{\alpha }  G_{\phi \gamma \gamma}
\right),
\label{eq:pm-to-gg}
\\
 \Gamma (\varphi_\pm \to f \bar{f} )
&
\simeq
\frac{g_{\pm ff}^2}{8 \pi } m_\pm,
&
 g_{\pm ff} 
&
= 
\frac{1}{X_+ - X_-}
 \left(\frac{X_{\pm} +  \beta^2}{ X_{\pm}^2 + \alpha^2 \beta^2
  }\right)^{-1/2}
\frac{X_\mp}{\alpha }
g_{\phi ff},
\label{eq:pm-to-ff}
\end{align}
where the fermion is assumed to be much lighter than the scalars.

Fixing the metric to a flat FRW, $ ds^2 = -dt^2 + a(t)^2 d\bd{x}^2$,
the homogeneous equations of motion of the diagonal fields 
can be written as
\begin{equation}
 \ddot{\varphi}_\pm + (3 H + \Gamma_\pm) \dot{\varphi}_\pm
 + m_\pm^2 \varphi_\pm 
+ U_{\mathrm{int}}'(\phi) \frac{\partial \phi}{\partial \varphi_\pm}
= 0,
\label{eq:B.7}
\end{equation}
where we have incorporated the decay of the scalars
through~$L_{\mathrm{c}}$ in the form of an effective friction
term $\Gamma_\pm \dot{\varphi}_\pm$.  
These equations can be rewritten in terms of the inflaton and axion
fields as
\begin{equation}\label{eq:fullEoMs}
\begin{split}
 \ddot{\sigma} &+ 
\left( 3 H + \frac{X_+ \Gamma_+ - X_- \Gamma_-}{X_+ - X_-} \right) \dot{\sigma}
+ \alpha \frac{ \Gamma_+ - \Gamma_-}{X_+ - X_-} \dot{\phi}
+ \frac{m_\sigma^2 \sigma - \alpha V'(\phi)}{1-\alpha^2}
= 0,
\\
 \ddot{\phi } &+ 
\left( 3 H + \frac{X_+ \Gamma_- - X_- \Gamma_+}{X_+ - X_-} \right) \dot{\phi}
+ \alpha \beta^2 \frac{ \Gamma_+ - \Gamma_-}{X_+ - X_-} \dot{\sigma}
+ \frac{V'(\phi) - \alpha m_\sigma^2 \sigma }{1-\alpha^2}
= 0.
\end{split}
\end{equation}
Apart from the terms involving the decay widths, these equations can also
be derived directly from the original Lagrangian~(\ref{eq:Lagrangian}),
and hence are exact in the limit $\Gamma_\pm \to 0$,
even when the axion mass depends on the temperature.
The description of the scalar decay in the form of friction terms
should be understood to be an effective one,
which could fail for non-perturbative decay processes, or if the
self-interaction~$U_{\mathrm{int}}$ is significant such that the scalar
fields cannot be interpreted as a collection of particles. 
We also remark that, with a temperature-dependent axion mass, 
$\varphi_{\pm}$ do not completely diagonalize the Lagrangian;
this may also yield corrections to the description of the decay in the
equations of motion, as well as to the decay widths
(such as those shown in (\ref{eq:pm-to-gg}) and (\ref{eq:pm-to-ff})).

\bibliographystyle{JHEP}
\bibliography{axion}

\end{document}